\documentclass[10pt,conference,letterpaper]{IEEEtran}

\ifCLASSOPTIONcompsoc
  \usepackage[nocompress]{cite}
\else
  \usepackage{cite}
\fi

\IEEEoverridecommandlockouts

\usepackage[english]{babel}
\usepackage[utf8]{inputenc}
\usepackage[T1]{fontenc}

\usepackage{amssymb}
\usepackage{amsmath}
\usepackage{graphicx}
\graphicspath{{imgs/}{figures/}{plots/}}
\usepackage[scaled]{beramono}
\usepackage[numbers]{natbib}
\usepackage{paralist}
\usepackage{textcomp}

\usepackage[inline]{enumitem}
\usepackage{algorithm}
\usepackage[noend]{algpseudocode}

\usepackage{array}
\usepackage{subfigure}
\usepackage{booktabs}
\usepackage{datetime}
\usepackage{caption}
\usepackage{etoolbox}
\usepackage[detect-all]{siunitx}
\usepackage{blkarray}

\usepackage{hyphenat}
\usepackage{xspace}
\newcommand\R{\mathbb{R}} 
\newcommand\F{\mathcal{F}} 
\newcommand\T{\mathcal{T}} 
\newcommand\Nfoi{\mathcal{N}_{\text{foi}}} 
\newcommand\G{\mathcal{G}} 
\newcommand{\B}[1]{\mathbf{#1}} 
\newcommand{\infunc}{A} 
\newcommand{\outfunc}{A^{\prime}} 
\newcommand{\func}{g} 
\newcommand{\funcother}{h} 
\newcommand{\leftover}{residual\xspace} 
\newcommand{\Leftover}{Residual\xspace} 
\newcommand{\hdev}{hDev} 
\newcommand{\vdev}{vDev} 
\newcommand{\hdevinc}{\hdev Inc} 
\newcommand{\vdevinc}{\vdev Inc} 
\newcommand{\1}[1]{\mathbf{1}_{\{#1\}}} 
\newcommand{\reward}{\mathit{reward}} 
\newcommand{\Ie}{I.e.,\xspace}
\newcommand{\ie}{, i.e.,\xspace} 
\newcommand{\eg}{, e.g.,\xspace} 
\newcommand{\Eg}{E.g.,\xspace} 
\newcommand{\DNCv}{v2.7\xspace} 
\newcommand{\FFMILPAv}{v1.0\xspace} 
\newcommand{\CPLEXv}{v20.1\xspace} 
\newcommand{\OpenJDKv}{12\xspace} 

\newtheorem{theorem}{Theorem}
\newtheorem{definition}{Definition}
\newtheorem{corollary}{Corollary}
\newtheorem{proof}{Proof}

\author{
	\IEEEauthorblockN{
		\hspace{85pt} Fabien Geyer\IEEEauthorrefmark{1}\IEEEauthorrefmark{2} \hspace{95pt}
		Alexander Scheffler\IEEEauthorrefmark{3} \hspace{10pt}
		Steffen Bondorf\IEEEauthorrefmark{3} \\[5pt]}
	\IEEEauthorblockA{
		\begin{minipage}{.26\textwidth}
			\centering
			\IEEEauthorrefmark{1}Technical University of Munich \\
			Munich, Germany
		\end{minipage}
		\begin{minipage}{.22\textwidth}
			\centering
			\IEEEauthorrefmark{2}Airbus Central R\&T \\
			Munich, Germany
		\end{minipage}
		\begin{minipage}{.45\textwidth}
			\centering
			\IEEEauthorrefmark{3}Faculty of Computer Science \\
			Ruhr University Bochum, Germany
	\end{minipage}}
}

\usepackage{nameref}
\usepackage[capitalize,noabbrev]{cleveref}

\usepackage{xcolor}

\usepackage{xspace}

\newcommand\MethodName{DeepFP\xspace}
\newcommand\MethodNameN[1]{DeepFP\textsubscript{#1}\xspace}

\newcommand\LUDBFF{FIFO\xspace} 
\newcommand\LUDBFFhFP{FIFO-$\mu$FP\xspace} 
\newcommand\LUDBFFFP{FIFO-FP\xspace} 

\newcommand\RNDFP[1]{RND\textsubscript{$#1$}\xspace}
\newcommand\RNDhFP[1]{$\mu$RND\textsubscript{$#1$}\xspace}

\usepackage{acro}
\DeclareAcronym{NC}{short=NC, short-indefinite = an, long=Network Calculus}
\DeclareAcronym{DNC}{short=DNC, long=Deterministic Network Calculus}
\DeclareAcronym{SNC}{short=SNC, short-indefinite = an, long=Stochastic Network Calculus}
\DeclareAcronym{RTC}{short=RTC, short-indefinite = an, long=Real-Time Calculus}
\DeclareAcronym{PBOO}{short=PBOO, long=Pay Bursts Only Once}
\DeclareAcronym{PMOO}{short=PMOO, long=Pay Multiplexing Only Once}
\DeclareAcronym{ML}{short=ML, short-indefinite = an, long=Machine Learning}
\DeclareAcronym{NN}{short=NN, short-indefinite = an, long=Neural Network}
\DeclareAcronym{GN}{short=GN, long=Graph Network}
\DeclareAcronym{GNN}{short=GNN, long=Graph Neural Network}
\DeclareAcronym{RL}{short=RL, short-indefinite = an, short-indefinite = an, long=reinforcement learning}
\DeclareAcronym{SL}{short=SL, short-indefinite = an, long=supervised learning}
\DeclareAcronym{FP}{short=FP, short-indefinite = an, long=Flow Prolongation}
\DeclareAcronym{AFDX}{short=AFDX, short-indefinite = an, long=Avionics Full-DupleX Ethernet}
\DeclareAcronym{TSN}{short=TSN, long=Time-Sentitive Networking}
\DeclareAcronym{FIFO}{short=FIFO, short-indefinite = a, long=First-In First-Out}
\DeclareAcronym{LUDB}{short=LUDB, short-indefinite = an, long=Least Upper Delay Bound}
\DeclareAcronym{TMA}{short=TMA, long=Tandem Matching Analysis}
\DeclareAcronym{GPU}{short=GPU, long=Graphics Processing Unit}
\DeclareAcronym{GGNN}{short=GGNN, long=Gated Graph Neural Networks}
\DeclareAcronym{GRU}{short=GRU, long=Gated Recurrent Unit}
\DeclareAcronym{FFNN}{short=FFNN, short-indefinite = an, long=Feed-Forward Neural Network}
\DeclareAcronym{UREX}{short=UREX, long=Under-appreciated Reward Exploration}
\DeclareAcronym{LP}{short=LP, short-indefinite = an, long=Linear Program}
\DeclareAcronym{MILP}{short=MILP, long=Mixed-Integer Linear Program}
\DeclareAcronym{FFLPA}{short=FF-LPA, short-indefinite = an, long=Feedforward Linear Programming Analysis}
\DeclareAcronym{FFMILPA}{short=FF-MILPA, short-indefinite = an, long=Feedforward Mixed-Integer Linear Programming Analysis}
\DeclareAcronym{DEBORAH}{short=DEBORAH, short-indefinite = a, long=DElay BOund RAting AlgoritHm}
\DeclareAcronym{NCorgDNC}{short=NCorg DNC, short-indefinite = an, long=NetworkCalculus.org Deterministic Network Calculator}
\DeclareAcronym{foi}{short=foi, short-indefinite = an, long=flow of interest} 
\DeclareAcronym{PyG}{short=PyG, long=pytorch-geometric}

\IfFileExists{./acro.sty}{\acsetup{patch/maketitle=false}}{}

\begin{document}

\IEEEtitleabstractindextext{%

\begin{abstract}
The derivation of upper bounds on data flows' worst-case traversal times is an important task in many application areas.
For accurate bounds, model simplifications should be avoided even in large networks.
\acf{NC} provides a modeling framework and different analyses for delay bounding.
We investigate the analysis of feedforward networks where all queues implement \acf{FIFO} service.
Correctly considering the effect of data flows onto each other under \ac{FIFO} is already a challenging task.
Yet, the fastest available \ac{NC} \ac{FIFO} analysis suffers from limitations resulting in unnecessarily loose bounds.
A feature called \acf{FP} has been shown to improve delay bound accuracy significantly.
Unfortunately, \ac{FP} needs to be executed within the \ac{NC} \ac{FIFO} analysis very often and each time it creates an exponentially growing set of alternative networks with prolongations.
\ac{FP} therefore does not scale and has been out of reach for the exhaustive analysis of large networks.
We introduce \MethodName, an approach to make \ac{FP} scale by predicting prolongations using machine learning.
In our evaluation, we show that \MethodName can improve results in \ac{FIFO} networks considerably.
Compared to the standard \ac{NC} \ac{FIFO} analysis, \MethodName reduces delay bounds by \SI{12.1}{\percent} on average at negligible additional computational cost.
\end{abstract}

}

\title{Network Calculus with Flow Prolongation --\\
A Feedforward FIFO Analysis enabled by ML}

\maketitle

\IEEEdisplaynontitleabstractindextext

%
\IEEEpeerreviewmaketitle


\acresetall
\section{Introduction}
\label{sec:introduction}

Modern, newly developed networked systems are often required to provide some kind of minimum performance level.
Applications in domains such as the automotive and avionics sector \cite{Geyer2016} as well as factory automation often crucially rely on this minimum performance.
Their main focus is then on one important network property: the worst-case traversal time\ie the end-to-end delay, of data communication.
To that end, safety-critical applications that are crucial for the entire system's certification need to formally prove guaranteed upper bounds on the end-to-end delay of data flows.
A second characteristic of modern networks in these domains is that they are not designed from scratch anymore but derived from IEEE Ethernet.
For example, network standards based on Ethernet such as \ac{AFDX} or IEEE \ac{TSN} are becoming prevalent.
These standards usually follow a simple queueing design, ``\ac{FIFO} per priority queue''.
In the \ac{NC} analysis' point of view, this is essentially a \ac{FIFO} system model -- a model already hard to analyze without introducing simplifying worst-case assumptions.
We even aim at improving the analysis in both dimensions, result accuracy and execution time, by adding the \ac{FP} feature as well as \ac{ML}-predictions steering our proposed \ac{NC} \ac{FIFO} analysis.
Our contribution can be applied to any system designed around \ac{FIFO} multiplexing and forwarding of data that can be modeled with \ac{NC}.
For example, existing works on \ac{NC} modeling of the specific schedulers used in \ac{AFDX} or \ac{TSN} networks.

\ac{NC} is a framework with modeling and analysis capabilities.
For best results\ie tight delay bounds, both parts should be developed in lockstep to prevent mismatches in their respective capabilities. 
Unfortunately, this has not always been the case.
Adding assumptions like \ac{FIFO} to a system model is naturally easy, yet tracking the property across an entire feedforward network to consider mutual impact of flows is not.
Other prominent network properties such as \ac{PBOO}~\cite{LeBoudec2001} and \ac{PMOO}~\cite{Schmitt2008b} stating that data flows do not exhibit stop-and-go behavior nor overtake each other multiple times when crossing a tandem of servers, have found their way into some \ac{NC} analyses eventually.
However, the important \ac{PMOO} property is not available in the non-\ac{FIFO}~\cite{Rizzo2005,Schmitt2011} analysis and not necessarily considered by the fastest available \ac{FIFO} analysis~\cite{Bisti2008,Bisti2012,Scheffler2021}.

If a network property cannot be considered by \iac{NC} tandem analysis, it is replaced by a worst-case assumption in the analysis' internal view.
Therefore, improvements to \ac{NC} tandem analyses tried to remove such overly pessimistic assumptions by adding analysis features that improve the computed delay bound.
A somewhat different feature was recently presented with \ac{FP}~\cite{Bondorf2017c}.
It actively converts the network model given to the \ac{NC} analysis to a more pessimistic one. 
This new view is explicitly derived such that a shortcoming of the \ac{NC} \ac{FIFO} analysis is circumvented while delay bounds remain valid. 
It has been shown that this model transformation can simultaneously help implementing the \ac{PMOO} property to a larger extent \cite{GeyerSchefflerBondorf_RTAS2021}.

\ac{FP} is conceptually straight-forward: assume flows take more hops than they actually do.
Nonetheless \ac{FP} is a powerful feature to add to \iac{NC} analysis that was adopted in \ac{SNC}~\cite{Nikolaus2020b}, too.
Finding the best prolongation of flows is prone to a combinatorial explosion.
On a tandem with $n$ hops and $m$ flows, there are $O(n^m)$ alternatives for path prolongation (including the alternative not to prolong any cross-flow).
Even with a deep understanding of the \ac{NC} analysis, the amount could not be reduced significantly to make \iac{FP} analysis scale~\cite{Bondorf2017c}.

Additionally, the fast \ac{FIFO} analysis derives multiple algebraic \ac{NC} terms, each bounding a single flow's delay.
The amount of terms grows exponentially with the network size and
none of them computes the tightest bound independent of flow parameters.
\Ie all need to be derived and solved~\cite{Bondorf2017a}, 
Thus, the analysis must compute a multitude of delay bounds to find the minimum among them~\cite{Bisti2008,Bisti2012}.
Adding the \ac{FP} feature to a \ac{FIFO} analysis for feedforward networks results in a hardly scalable analysis. 

\begin{figure}[t]
	\centering
	\includegraphics[width=\columnwidth]{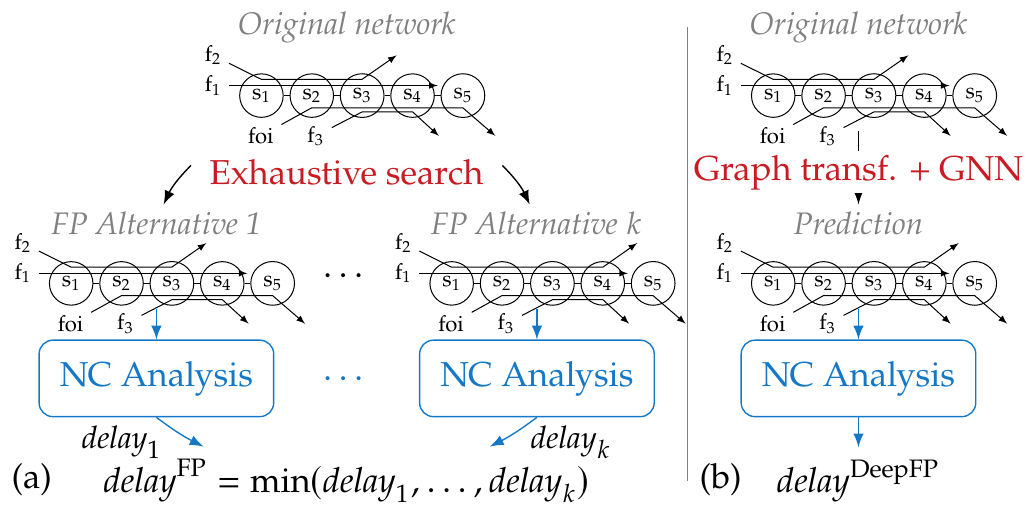}
	\caption{Comparison between the (a) exhaustive \ac{FP} with $O(n^m)$ \ac{NC} analyses and (b) \MethodName with one prediction.}
	\label{fig:approach}
\end{figure}

In this article, we present \iac{ML} approach to overcome exhaustive searches in the algebraic \ac{NC} \ac{FIFO} analysis with the \ac{FP} feature.
Put simple, we have trained a \ac{GNN} to predict the best choices for the analysis' term creation.
This allows the analysis to scale, and can also be extended to multiple predictions per choice.
We base our contribution on some previous work~\cite{Geyer2019a,Geyer2020,Geyer2020b,GeyerSchefflerBondorf_RTAS2021}, presenting here the \textit{\MethodName} analysis illustrated in \cref{fig:approach}.

Via numerical evaluations, we show that \MethodName is an efficient and scalable method that reduces \ac{NC} \ac{FIFO} delay bound by an average of \SI{12.1}{\percent} on our test networks.
We show that \MethodName scales to networks with up to 500 flows without compromising on tightness, whereas other methods fail to finish within a deadline of \SI{1}{\hour} and / or a memory limitation of \SI{5}{GB}.
Compared to previous work~\cite{Geyer2019a,Geyer2020,Geyer2020b,GeyerSchefflerBondorf_RTAS2021}, we train \MethodName using a \ac{RL} approach and show that it even outperforms an expert heuristic for \ac{FP} on small networks.

This article is organized as follows:
\Cref{sec:relatedwork} presents related work and \cref{sec:netcal_analysis_short} gives an overview on \ac{NC}.
\Cref{sec:fp4ludb} shows how \ac{FP} improves the \ac{NC} \ac{FIFO} analysis and the challenges it imposes.
\Cref{sec:contribution} contributes our \MethodName, making \ac{FP} applicable in the network analysis.
\Cref{sec:numerical_evaluation} evaluates \MethodName
and \cref{sec:conclusion} concludes the article.


\section{Related Work}
\label{sec:relatedwork}

\subsubsection*{\acl{NC} and \acl{RTC}}

\ac{NC} creates a purely descriptive model of a network of queueing locations and data flows (see \cref{sec:netcal}).
The \ac{NC} analysis then computes a bound on the worst-case delay for a certain flow, the so-called \ac{foi} (see \cref{sec:netcal_analysis}).
A variant of \ac{NC} that focuses on (embedded) real-time systems was established by the \acf{RTC}~\cite{Thiele2000}. 
Equivalence between the slightly differing resource descriptions has been proven in~\cite{Bouillard2009}.
What remains is the difference in modeling of the ``network'' and the analysis thereof.
\ac{RTC} models networks of components such as the Greedy Processing Component (GPC)~\cite{Guan2013,Tang2017}.
Each component represents a macro\ie a fixed sequence of algebraic \ac{NC} operations to apply to its input.
Thus, the model already encodes the analysis.
Moreover, this component-based modeling approach mostly restricts the analysis to strict priority multiplexing, yet, efforts to incorporate the \ac{PBOO} property and the \ac{PMOO} property can be found in the literature \cite{Lampka2016,Lampka2017,Tang2019}.
We, in contrast, aim for a model-independent improvement of the automatic derivation of a valid order of \ac{NC} operations -- the process called \ac{NC} analysis -- for feedforward networks of \ac{FIFO} queues.
First results on this topic in NC~\cite{LeBoudec2001,Fidler2004} were refined to the \ac{LUDB} analysis~\cite{Bisti2008,Bisti2012} that we use as \textit{the (algebraic) \ac{NC} \ac{FIFO} analysis} in our work.
Later, an entirely different analysis approach was proposed in~\cite{Bouillard2012,Bouillard2015}. 
It converts the \ac{NC} entire feedforward network model to a single optimization problem, optimizing for the \ac{foi}'s delay bound.
The tight \ac{MILP} formulation introduces forbiddingly large computational effort such that it was augmented with a less tight \ac{LP} formulation -- we call these analyses \ac{FFMILPA} and \ac{FFLPA}.
Current efforts in this stream of research further tune the \ac{FFLPA}'s tradeoff between delay bound tightness and computational effort by adding constraints derived with algebraic \ac{NC}~\cite{Bouillard2022}.
We numerically compare \MethodName with \ac{FFLPA} and \ac{FFMILPA} in \cref{sec:numerical_evaluation}.

\subsubsection*{\acl{FP} in \acl{NC}}

\Ac{FP} was mentioned in \cite{Schmitt2007} to be used for the purpose of complexity reduction of \iac{NC} analysis.
This is achieved if a flow can be prolonged to share the same path as another flow, allowing both to be aggregated within the analysis.
In contrast, we pair \ac{FP} with the (\ac{LUDB}) \ac{NC} \ac{FIFO} analysis for feedforward networks~\cite{Scheffler2021} to counteract its main tightness-compromising problem, thus considerably improving delay bounds.
There has been one previous mention of \ac{FP} in \ac{FIFO} networks: 
\cite{Bisti2008} briefly shares the observation that, if prolonged, a cross-flow can be aggregated with the \ac{foi} -- which is independent of the problem we tackle in this article.
The observation of \cite{Bisti2008} can thus be combined with our contribution.
Yet, as it only applies to at most one single point in the analysis, we will focus our article on investigating the \ac{FP}-improvement throughout the entire feedforward \ac{NC} \ac{FIFO} analysis and will leave this smaller feature's investigation to future work.
Prolonging at the front may also be possible in the arbitrary multiplexing \ac{PMOO} analysis~\cite{Bondorf2020}.

\subsubsection*{\aclp{GNN}}
\label{sec:relatedwork:gnn}

\acp{GNN} were first introduced in~\cite{Gori2005} and~\cite{Battaglia2018} presents a framework that formalizes many concepts applied in \acp{GNN} in a unified way.
\acp{GNN} were already proposed as an efficient method for replacing exhaustive searches or similar NP-hard problems such as the traveling salesman problem~\cite{Prates2019}.
A recent survey~\cite{Wang2020} about existing applications of \ac{ML} to formal verification shows 
that this combination can accelerate, for instance, theorem proving, model-checking, Boolean satisfiability (SAT) and satisfiability modulo theories (SMT) problems.
As we show, \ac{NC} has been combined with other methods, too.

\acp{GNN} were first introduced in~\cite{Gori2005,Scarselli2009}.
a concept subsequently refined in recent works.
Gated Graph Neural Networks (G\acp{GNN})~\cite{Li2016a} extended this architecture with modern practices by using Gated Recurrent Unit (GRU) memory units~\cite{Cho2014b}.
Message-passing neural network were introduced in~\cite{Gilmer2017}, with the goal of unifying various \ac{GNN} and graph convolutional concepts.
Finally,~\cite{Battaglia2018} introduced the \ac{GN} framework, a unified formalization of many concepts applied in \acp{GNN}.

Concrete examples are:
\cite{Li2018} that tackles the challenge of solving combinatorial optimization problems using Graph Convolutional Networks.
\cite{Selsam2019} addresses constraint satisfaction problems (CSP) and particularly SAT problems using \acp{GNN}, showing that \acp{GNN} can be used for such problems, hence, an application to \ac{NC} is also feasible.
\cite{Palm2018} used Recurrent Relational Networks to solve Sudoku problems, and \cite{Prates2019} used \acp{GNN} to solve the traveling salesman problem.
Finally, recent work in \cite{Davies2021} applies \acp{GNN} to mathematical theorem proving.

For computer networks, \acp{GNN} have lately been applied to different non-\acp{NC} performance evaluations of networks~\cite{Geyer2018e,Rusek2020,Suzuki2020}.
\Eg \cite{Mai2021} used \acp{GNN} for predicting the feasibility of scheduling configurations in Ethernet networks.
Surveys on \ac{GNN} applications to communication networks and network optimization can be found in \cite{Vesselinova2020,Jiang2022}.

\subsubsection*{\acl{NC} and \aclp{GNN}}

DeepTMA was proposed in~\cite{Geyer2019a} as a framework where \acp{GNN} are used for predicting the best contention model whenever the analysis is faced with alternatives.
\MethodName and DeepTMA are closely related in spirit: both methods use a graph transformation and a \ac{GNN} to replace a computationally expensive exhaustive search.
DeepTMA, as it name suggests, does so within the \ac{TMA}~\cite{Bondorf2017a}. 
\ac{TMA} does not consider -- and for that matter nor trace -- the \ac{FIFO} property but instead replaces it with a worst-case assumption whenever flows multiplex in a queue.
That results in a considerably less complex approach that was shown to scale to large feedforward networks with up to \num{14000} flows~\cite{Geyer2020b}.
But as shown in~\cite{Scheffler2021}, \ac{TMA}-derived delay bounds are overly pessimistic in \ac{FIFO} networks.
In contrast, \MethodName pairs the \ac{NC} \ac{FIFO} analysis, including the \ac{FP} feature, with a \ac{GNN}.
This requires us, among other challenges, to design a new graph transformation and to connect all the parts efficiently.

Various other works using \acp{GNN} for achieving better delay bounds or optimizing network configurations have been proposed.
\cite{Geyer2018d} proposed to predict the delay bound computed by different \ac{NC} analyses by using \acp{GNN}.
This information can be used to select the analysis that will most likely deliver the best delay bound\ie unlike our \MethodName or DeepTMA, the \ac{GNN} predictions do not impacted the proceeding of the analysis itself.
Finally,~\cite{Mai2021} recently used \acp{GNN} for predicting the feasibility of scheduling configurations in Ethernet networks.


\section{Network Calculus Analyses}
\label{sec:netcal_analysis_short}


\subsection{\acl{NC} System Model}
\label{sec:netcal}

\ac{NC} models \cite{LeBoudec2001,Bouillard2018,Chang2000} a network as a directed graph of connected queueing locations, the so called server graph.
A server offers a buffer to queue incoming demand (the data) and a service resource (forwarding of data).
Data is injected into the network by flows.
We assume unicast flows with a single source server and a single sink server as well as a fixed route between them.
Flows' forwarding demand is characterized by functions cumulatively counting their data,
\begin{equation}
\F^{+}_{0}\!=\!\left\{ \func:\R^{+}\rightarrow\R^{+}\,|\,\func(0)\!=\!0,\;\forall s\le t\,:\,\func(t)\!\geq\!\func(s)\right\} .
\end{equation}

Let functions $\infunc(t)\in\F^{+}_{0}$ denote a flow's data put into a server and let $\outfunc(t)\in\F^{+}_{0}$ be the flow's data put out of, both in the time interval $[0,t)$.
We require the input/output relation to preserve causality by $\forall t\in\R^{+}:\;\infunc(t)\ge \outfunc(t)$.

\ac{NC} refines this model by using bounding functions (called curves).
They are defined independent of the start of observation and instead by the duration of the observation. 
By convention, let curves be in set $\F_{0}$ that simply extends the definition of $\F^{+}_{0}$ by $\forall t\le 0 : \func(t)\!=\!0$.

\begin{definition}[Arrival Curve]
\label{def:Arrival-Curve}
Given a flow with input $\infunc$, 
a function $\alpha\in\F_{0}$ is an arrival curve for $\infunc$ iff
	\begin{equation}
	\forall\,0\leq d\le t\,:\,\infunc(t)-\infunc(t-d)\leq\alpha(d).
	\end{equation}
\end{definition}

Complementing data arrivals, the service offered by a server~$s$ is modeled with a lower bounding curve:

\begin{definition}[Service Curve]
If the service by server $s$ for a given input $\infunc$ results in an output $\outfunc$, then $s$ offers a service curve $\beta \in \F_0$ iff
\begin{equation}
	\forall t : \outfunc(t) \geq \inf_{0 \leq d \leq t} \{ \infunc(t - d) + \beta(d) \}.
\end{equation}
\end{definition}

In this article, we restrict the set of curves to affine curves (the only type that can be used with the \ac{FIFO} analysis).
These curves are suitable to model token-bucket constrained data flows
$\gamma_{r,b}\!:\R^{+}\rightarrow\R^{+}\,|\,\gamma_{r,b}\left(0\right)\!=0,\underset{d>0}{\forall}\gamma_{r,b}(d)\!=b+r\cdot d$,
$r,b \ge 0,$
where $b$ upper bounds the worst-case burstiness and $r$ the arrival rate.
Secondly, rate-latency service can be modeled by affine curves
$
\beta_{R,T}:\R^{+}\rightarrow\R^{+}\,|\,\beta_{R,T}\left(d\right)=\max\{0,R\cdot(d-T)\}$, 
$T \ge 0,\,R > 0
$
where $T$ upper bounds the service latency and $R$ lower bounds the forwarding rate.

\subsection{Algebraic \acl{NC} Analysis}
\label{sec:netcal_analysis}

An \ac{NC} analysis aims to compute a bound on the worst-case delay that a specific \ac{foi} experiences on its path.
The algebraic analysis \cite{LeBoudec2001,Bouillard2018,Chang2000} does so by deriving a (min,plus)-algebraic term that bounds the delay.
Service curves on a path are shared by all flows crossing the respective server  
yet an arrival curve is only known at the respective flow's source server.
To derive the \ac{foi}'s end-to-end delay bound from such a model, the \ac{NC} analysis relies on (min,plus)-algebraic curve manipulations.

\begin{definition}[\ac{NC} Operations]
\label{def:MinPlusOperations}
The (min,plus)-algebraic aggregation, convolution and deconvolution
of two functions $\func,\funcother\in\F_{0}$ are defined as
\begin{eqnarray}
\text{aggregation:} &  & \hspace{-6.5mm}\left(\func+\funcother\right)\left(d\right) = \func\left(d\right)+\funcother\left(d\right)\!,\\
\text{convolution:} &  & \hspace{-6.5mm}\left(\func\otimes \funcother\right)(d) = \hspace{-1.5mm}{\displaystyle \inf_{0\leq u\leq d}}\hspace{-0.75mm}\left\{ \func(d-u)+\funcother(u)\right\} \hspace{-0.75mm},\\
 \hspace{-5.5mm}\text{deconvolution:} &  & \hspace{-6.5mm}\left(\func \oslash \funcother\right)(d) = \sup_{u\geq0}\left\{ \func(d+u)-\funcother(u)\right\}\!.
\end{eqnarray}
\end{definition}

Aggregation of arrival curves creates a single arrival curve for their multiplex.
With convolution, a tandem of servers can be treated as a single server providing a single service curve. 
Deconvolution allows to compute an arrival curve bounding a flow's (or flow aggregate's) $A^{\prime}(t)$ after a server. 
Backlog and delay can be bounded as follows:

\begin{theorem}[Performance Bounds]
\label{thm:Performance-Bounds} 
Consider a server $s$ that offers a service curve $\beta$. Assume
a flow $f$ with arrival curve $\alpha$ traverses the server. Then
we obtain the following performance bounds for $f, \forall t\in\R^{+}$:
\begin{eqnarray}
\mathcal{\textrm{output:}} & & \hspace{-5mm} \alpha^{\prime} (t) = \left(\alpha \oslash \beta\right) (t) \\
\mathcal{\textrm{backlog:}} & & \hspace{-5mm} B\left(t\right)\leq\left(\alpha\oslash\beta\right)(0) = \vdev(\alpha, \beta)\\
\mathcal{\textrm{delay:}} & & \hspace{-5mm} D\left(t\right)\leq\inf\left\{ d\geq0\;|\;\left(\alpha\oslash\beta\right)(-d)\leq0\right\} \nonumber \\
& & \hspace{-5mm}  = \hdev(\alpha, \beta)
\end{eqnarray}
\end{theorem}

Given curves $\beta$ lower bounding available forwarding service and $\alpha$ upper bounding arriving data, 
\ac{NC} can compute lower bounds on \iac{foi}'s \leftover service.

\begin{theorem}[\Leftover Service Curves for \ac{FIFO} Multiplexing]
\label{thm:Left-Over}
Let flow $f$ have arrival curve $\alpha$.
The guaranteed \leftover service at a \ac{FIFO} server~$s$ serving $f$ is bounded by the infinite set of  service curves \cite[Theorem 4]{Cruz1998} 
\begin{eqnarray} 
	\hspace{2mm} \beta_{\theta,\alpha}(t) & = & [\beta(t) - \alpha(t-\theta)]^{\uparrow} \cdot \1{t > \theta} \nonumber \\
	& \;=: & \beta\ominus_{\theta}\alpha \label{eq:fifo_mux_lo}, \; \forall \theta \geq 0
\end{eqnarray}
where
$\1{\text{condition}} $ denotes the test function (1 if the condition is true, 0 otherwise) and 
$[\func(x)]^{\uparrow} = \sup_{0\leq z \leq x} \func(z)$
is the non-decreasing closure
of $\func(x)$, defined on positive real values $\in \R^+_0$.
\end{theorem}

The free parameter $\theta$ captures the potential impact of \ac{FIFO} multiplexing without the need to already know other flows at $s$.
It defines an infinite set of (valid) \leftover service curves given $f$'s presence. 
The best $\theta$ setting depends on the \ac{foi} arrival curve served by $\beta_{\theta,\alpha}$ as well as the performance characteristic to be bounded -- \ac{foi} delay, backlog or output. 
Finding the best $\theta$ requires the \ac{foi}'s arrival curve and was extensively investigated in~\cite{Lenzini2005,Bisti2010}.

\subsection{The \ac{NC} \ac{FIFO} Analysis: \acl{LUDB}}
\label{sec:netcal:ludb}

The (algebraic) \ac{NC} \ac{FIFO} analysis makes heavy use of \cref{thm:Left-Over},
imposing the challenge to set all interdependent $\theta$ parameters for a best possible bound.
The \ac{LUDB} analysis presented in the seminal work of~\cite{Bisti2008,Bisti2012} converts the algebraic \ac{NC} term with its $\theta$ parameters to \iac{LP} formulation.
The transformation is restricted to token-bucket service and rate-latency arrivals.
Initially, \ac{LUDB} focused on bounding the \ac{foi}'s delay\ie providing a tandem analysis, while latest work~\cite{Scheffler2021} embedded it into the \ac{NC} feedforward analysis resulting in the analysis \ac{LUDB}-FF, thus increasing the need to optimize for output bounding.
For simplicity and independence of the actual implementation, we will use \textit{\ac{LUDB}-FF} and \textit{\ac{FIFO} analysis} interchangeably.

The more important challenge for our article is, however, the implementation of the \ac{PMOO} principle that is essential for achieving tight bounds.
\ac{LUDB} does not necessarily achieve a full implementation of the \ac{PMOO} principle.
Success in doing so depends on the nesting of flows on the analyzed tandem.
A tandem is called \textit{nested} if any two flows have disjunct paths or one flow is completely included in the path of the other flow. 
For example, in \cref{fig:example_network:nc}, $f_2$ is nested into $f_1$
but neither is nested into the \ac{foi}.
\ac{LUDB} must cut a non-nested tandem into a sequence of nested sub-tandems.
These cuts deprive it of its end-to-end view on the tandem and compromise the \ac{PMOO} property.
Moreover, sequences of nested tandems can be created by alternative cut sets, each of which needs to be analyzed to find the best\ie least upper, delay bound.
As we will show in \cref{sec:fp4ludb}, we can employ \ac{FP} to reduce the number of cut sets.

\subsection{\acl{FP} Fundamentals}
\label{sec:fp_background}

\ac{FP} was designed as an add-on feature to mitigate a problem in the feedforward analysis when bounding the arrivals of cross-flows that interfere with the \ac{foi}~\cite{Bondorf2017c}.
\ac{FP} is nonetheless independent of any purpose, as its properties show:

\begin{corollary}[Delay Increase due to \ac{FP}]
\label{cor:delay_inc_FP}
  Assume a tandem of servers $\T$ defined by \iac{foi}'s path.
  A prolongation of cross-flows cannot decrease the end-to-end delay experienced by the \ac{foi}.
\end{corollary}
\begin{proof}
  Let the \ac{foi} be called with $f_1$ and wlog assume a single cross-flow $f_2$.
  Prolong $f_2$ over one additional server $s$ on $\T$ to create the tandem $\T_{\text{FP}}$.
  Compared to $\T$, $s$ in $\T_{\text{FP}}$ multiplexes incoming data of $f_2$ with data of $f_1$ in its queue.
  $s$ either forwards the data of $f_2$ after $f_1$, causing no increase of $f_1$'s delay on $\T_{\text{FP}}$, 
  or it forwards at least parts of the data of $f_2$ before $f_1$\eg due to \ac{FIFO} queueing, causing an additional queuing delay to $f_1$.
\end{proof}

\Cref{cor:delay_inc_FP} shows that \ac{FP} is a conservative transformation adding pessimism to the network model that increases the \ac{foi}'s delay. 
For delay bounds, it holds that:

\begin{corollary}[\ac{FP} Delay Bound Validity]
\label{cor:delay_bound_FP}
  Assume a tandem of servers $\T$ defined by \iac{foi}'s path.
  Let $\T_{\text{FP}}$ be derived from $\T$ by \ac{FP}.
  Then, the bound on the \ac{foi}'s worst-case delay in $\T_{\text{FP}}$ is a bound on the \ac{foi}'s delay on $\T$.
\end{corollary}
\begin{proof}
  Per \cref{cor:delay_inc_FP}, we know that the \ac{foi}'s end-to-end delay will not decrease by FP.
  Thus, the tight delay bound in $\T_{\text{FP}}$ will exceed the tight delay bound on $\T$ 
  and any potentially untight bound derived for $\T_{\text{FP}}$ is also a bound on the \ac{foi}'s delay on $\T$.
\end{proof}


\section{A Feedforward \ac{NC} \ac{FIFO} Analysis with \ac{FP}}
\label{sec:fp4ludb}

In this Section, we address the question of how \ac{FP}, 
despite being a network transformation that increases the actual delay of flows,
can improve the NC-derived worst-case delay bound for a \ac{foi} in the \ac{FIFO} analysis.

\subsection{\ac{FP} in the Feedforward \ac{NC} \ac{FIFO} Analysis}
\label{sec:fp4ludb:ff_fp}

The algebraic \ac{NC} feedforward analysis that we aim to improve with \ac{FP} is compositional.
It decomposes a feedforward network into a sequence of tandems to analyze -- starting with the \ac{foi}'s path where its delay is bounded, followed by backtracking cross-flows whose output is bounded.
An alternative way to view the problem is to match tandems onto the feedforward network -- giving name to the \ac{TMA}~\cite{Bondorf2017a} and DeepTMA~\cite{Geyer2019a}.
The network analysis procedure itself is independent of any queueing assumption, allowing to bring \ac{FP} of \cite{Bondorf2017c} to the \ac{FIFO} analysis.

\begin{figure}[h]
	\centering
	\subfigure[Example tandem network shown in \cref{fig:approach}]{\label{fig:example_network:nc}
		\includegraphics[width=.6\columnwidth,page=1]{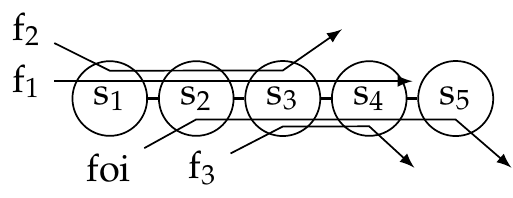}}
	\subfigure[Indication of all its potential flow prolongation alternatives]{\label{fig:example_network:nc_fps}
		\includegraphics[width=.6\columnwidth,page=1]{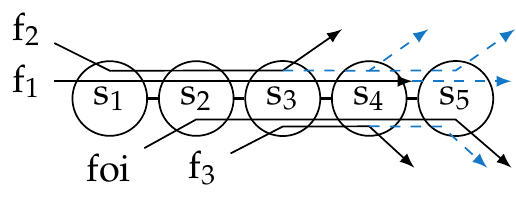}}
	\subfigure[A beneficial prolongation]{\label{fig:example_network:nc:single_fp}
		\includegraphics[width=.6\columnwidth,page=1]{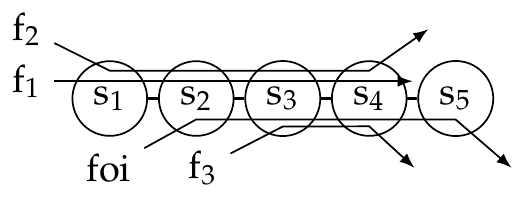}}
	\caption{Sample network with prolongations and a specific \ac{FP} alternative we will evalaute as an example}
	\label{fig:example_network:nc:all}
\end{figure}

During the feedforward analysis, \ac{FP} can improve the derived bounds in two distinct ways:
\begin{enumerate}[label=(I\arabic*)]
	\item on each tandem, it can reduce the amount of cuts required in the \ac{FIFO} analysis, \label{itm:fp_improv:cuts}
	\item when backtracking cross-flows, it can allow for aggregate computation of output bounds. \label{itm:fp_improv:aggr}
\end{enumerate}
While these two improvements are distinct, they are not necessarily isolated from each other as we will illustrate on the small sample network shown in \cref{fig:example_network:nc:all}.

Take the sample tandem in \cref{fig:example_network:nc}, where bounding the arrivals of data flows $f_1$ and $f_2$ is required at their first location of interference with the \ac{foi}, server $s_2$.
Independent of the \ac{FIFO} multiplexing assumption, the \ac{NC} feedforward analysis may suffer\footnote{If the \ac{NC} analysis does not trace the \ac{FIFO} property, it was shown that aggregate bounding of flows is generally preferable~\cite{Bondorf2015b}. This also holds for \ac{FIFO} systems. It was later shown that enforcing a segregated view can also be beneficial due to analysis drawbacks~\cite{Bondorf2018}, yet, it remains unproven if this can potentially improve bounds in the analysis of \ac{FIFO} systems.} from the so-called segregation effect~\cite{Bondorf2017a}: 
both flows assume to only receive service after the respective other flow was forwarded by server $s_1$ -- an unattainable pessimistic forwarding scenario in the analysis' internal view on the network.
With \ac{FP}, it is possible to steer the analysis such that it does not have to apply this pessimism.
By prolonging flow $f_2$ to also cross $s_4$, we match its path with $f_1$ (see \cref{fig:example_network:nc_fps}) and allow both flows' output from $s_1$ to be bounded in aggregate.
Note, that this adds interference to the \ac{foi} at $s_4$ and may therefore not be beneficial for its delay bound after all.
Thus, usually all alternative prolongations of flows as shown by the dashed lines in \cref{fig:example_network:nc_fps} must be tested.
Although only aiming at improvement \ref{itm:fp_improv:aggr}, the approach was shown to not scale~\cite{Bondorf2017c}.

The dominant problem that causes a loss of tightness in the \ac{NC} \ac{FIFO} analysis is the lack of the \ac{PMOO} property, caused by the cuts mentioned in improvement \ref{itm:fp_improv:cuts} above.
In general, the property is achieved by analyses that first create an end-to-end view on the tandem under analysis.
Algebraic \ac{FIFO} analysis cannot achieve this for non-nested interference patterns (see \cref{sec:netcal:ludb} and \cref{fig:example_network:nc}, servers $s_2$ to $s_4$ where cross-flow paths overlap).
It can only analyze nested tandems in an end-to-end fashion allowing for full implementation of the \ac{PMOO} property.
\ac{FP} allows for transforming a non-nested tandem into a nested one.
See \cref{fig:example_network:nc:single_fp} where the path of $f_2$ was prolonged over $s_4$, causing $f_3$ to become nested into $f_2$ instead of creating a non-nested interference pattern with it.

\subsubsection*{Impact Evaluation on Sample Network in \Cref{fig:example_network:nc:all}}
As mentioned above and illustrated by \cref{fig:example_network:nc:all}, the two improvements that can be attained with \ac{FP} do not necessarily occur in isolation.
We provide a detailed evaluation of the impact of the proposed prolongation in our example.

\begin{figure}[h!]
	\centering
		\subfigure[Cutting left of server $s_3$]{\label{fig:example_network:ludb_ff_cuts:left}
			\includegraphics[width=.66\columnwidth]{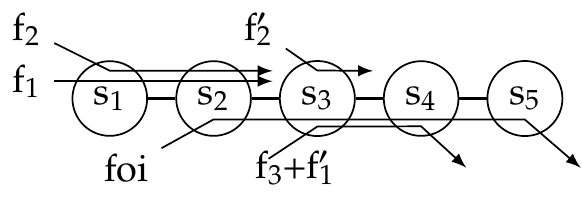}
		}%
	\\
		\subfigure[Cutting right of server $s_3$]{\label{fig:example_network:ludb_ff_cuts_right}
			\includegraphics[width=.66\columnwidth]{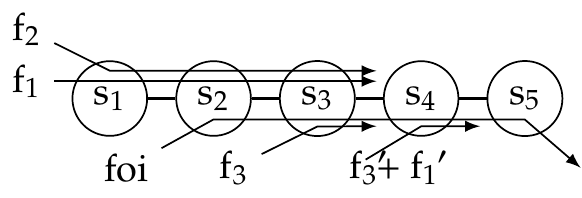}
		}
	\caption{Sample network in \cref{fig:example_network:nc} analyzed with \LUDBFF requires find the best among two possible cut locations}
	\label{fig:example_network:ludb_ff_cuts}
\end{figure}

To apply the \ac{FIFO} analysis to non-nested tandems, such a tandem is cut into a sequence of sub-tandems that all have nested interference patterns only.
In \cref{fig:example_network:nc}, the tandem can be cut before or after server $s_3$ as shown in \cref{fig:example_network:ludb_ff_cuts}.
Either alternative has the very same drawback: a cross-flow is cut, too, depriving the analysis of its end-to-end view on said flow.
\ac{PMOO} is not achieved and bounds become untight.

Let server $s_i$ provide service $\beta_i$ and let flow $f_j$ put $\alpha_j$ data into the network. 
Further denote with $\alpha^{s}$ some arrival curve at server $s$.
The respective (min,plus)-algebraic terms bounding the \ac{foi}'s end-to-end \leftover service curve are:
\begin{eqnarray}
& & \hspace{-25mm} \beta_{\Theta_\text{I},\text{i},\text{left}} \; = \; (((\beta_3 \ominus_{\theta_5} \alpha^{s_3}_2) \otimes \beta_4) \ominus_{\theta_6} (\alpha_3 + \alpha^{s_3}_1)) \nonumber \\
& & \hspace{5mm} \otimes (\beta_2 \ominus_{\theta_7} ((\alpha_1 + \alpha_2 )\oslash \beta_1)) \otimes \beta_5 \label{eqn:beta_lo:cut_left} \\
 \hspace{-5mm}\text{with}\hspace{5mm} \alpha^{s_3}_1 & = & \alpha_1 \oslash ((( \beta_2 \ominus_{\theta_1} \alpha_{\text{foi}}) \otimes \beta_1 ) \ominus_{\theta_2} \alpha_2) \nonumber \\ 
\alpha^{s_3}_2 & = & \alpha_2 \oslash ((( \beta_2 \ominus_{\theta_3} \alpha_{\text{foi}}) \otimes \beta_1) \ominus_{\theta_4} \alpha_1) \nonumber
\end{eqnarray}%
for the cut left of $s_3$ and $\Theta_\text{I} = \{\theta_1,\ldots,\theta_7\}$ and for the cut right to it with $\Theta_\text{II} = \{\theta_8,\ldots,\theta_{16}\}$ to later optimize:%
\begin{eqnarray}
& & \hspace{-20mm} \beta_{\Theta_\text{II},\text{i},\text{right}} \; = \; (((\beta_3 \ominus_{\theta_{14}} \alpha_3) \otimes \beta_2) \ominus_{\theta_{15}} ((\alpha_1 + \alpha_2) \oslash \beta_1))  \nonumber \\
& & \hspace{27.5mm} \otimes  (\beta_4 \ominus_{\theta_{16}} \alpha^{s_4}) \otimes \beta_5  \label{eqn:beta_lo:cut_right} \\
\text{with}\hspace{2mm}\alpha^{s_4} & = & (\alpha_3 + (\alpha_1 \oslash (((\beta_2 \ominus_{\theta_8} \alpha_{\text{foi}}) \otimes \beta_1) \ominus_{\theta_9} \alpha_2))) \nonumber \\
& & \hspace{39mm}\oslash (\beta_3 \ominus_{\theta_{10}} \alpha^{s_3}) \nonumber \\
\alpha^{s_3} & = & ((\alpha_2 \oslash (\beta_1 \ominus_{\theta_{11}} \alpha_1)) + \alpha_{\text{foi}}) \nonumber \\ 
& & \hspace{18mm}\oslash (\beta_2 \ominus_{\theta_{12}} (\alpha_1 \oslash (\beta_1 \ominus_{\theta_{13}} \alpha_2))) \nonumber
\end{eqnarray}

Note\eg in $\alpha^{s_3}$ the segregation of flows at $s_1$ by the simultaneous occurrence of $\alpha_1 \oslash (\beta_1 \ominus_\theta \alpha_2)$ and $\alpha_2 \oslash (\beta_1 \ominus_\theta \alpha_1)$ in the term that will be used for the right sub-tandem.
Moreover note, that the sub-tandem to the left of $s_3$ can, in contrast, apply aggregate bounding thanks to flows' common last server on it. 
The analysis thus computes $(\alpha_1 + \alpha_2) \oslash \beta_1$.

With the \ac{FP} alternative shown in \cref{fig:example_network:nc:single_fp}, neither segregation nor cutting is required and the \leftover service curve is significantly less complex with $\Theta_\text{III} = \{\theta_a,\theta_b\}$:
\begin{equation}\label{eqn:beta_lo:fp_alt}
\beta_{\Theta_\text{III},\text{iii}} = ((((\beta_3 \otimes \beta_4)\ominus_{\theta_a} \alpha_3)\otimes\beta_2) \ominus_{\theta_b} ((\alpha_1 + \alpha_2) \oslash \beta_1)) \otimes \beta_5
\end{equation}

We have tested this instantiation of \ac{FP} for different curve settings and w.r.t. its impact on either bounding the \ac{foi} delay or output. 
Service curves were set to $\beta_{R=40,T}$ and arrival curves to $\gamma_{r = u \cdot 10, b}$ where $u$ denotes the utilization $\frac{4r}{R}$ at the server that always sees four flows, $s_3$.
Latencies $T$ as well as bursts $b$ were set to 0, 0.1 or 10.
Note, that our setting guarantees for finite bounds.

For quantification of the improvement, we compute the reduction of the respective bound's variable part relative to the \LUDBFF bound in the original network.
That is, the end-to-end delay bound cannot be decreased below the sum of latencies on the analyzed \ac{foi}'s path.
We incur a fixed latency of $\sum_{i=2}^{5} T_i = 4T$ in our sample network that we subtract from the bounds computed for the original and the \ac{FP} network. 
We call the remaining variable part of the delay bound \textit{latency increase} $\hdevinc$.
Its improvement is defined as follows:
\begin{eqnarray}
& & \hspace{-7.5mm} \frac{\mathit{delay}^\text{FIFO} - \mathit{delay}^\text{FIFO-FP\textsubscript{III}}}{\mathit{delay}^\text{FIFO}} \nonumber \\
& &\hspace{-7.5mm} = \frac{\hdev(\alpha_\text{foi},\beta_{\Theta_{\text{I}\wedge\text{II}},\text{i}}) - \hdev(\alpha_\text{foi},\beta_{\Theta_\text{III},\text{iii}})}{\mathit{delay}^\text{FIFO}} \nonumber \\
& &\hspace{-7.5mm}  = \frac{(\hdev(\alpha_\text{foi},\beta_{\Theta_{\text{I}\wedge\text{II}},\text{i}}) - 4T) - (\hdev(\alpha_\text{foi},\beta_{\Theta_\text{III},\text{iii}}) - 4T)}{\mathit{delay}^\text{FIFO}} \nonumber \\
& &\hspace{-7.5mm} = \frac{\hdevinc(\alpha_\text{foi},\beta_{\Theta_{\text{I}\wedge\text{II}},\text{i}}) - \hdevinc(\alpha_\text{foi},\beta_{\Theta_\text{III},\text{iii}})}{\mathit{delay}^\text{FIFO}}\label{eq:latency_inc}
\end{eqnarray}
where $\hdev(\alpha_\text{foi},\beta_{\Theta_{\text{I}\wedge\text{II}},\text{i}}) = \hdev(\alpha_\text{foi},\beta_{\Theta_\text{I},\text{i},\text{left}}) \wedge \hdev(\alpha_\text{foi},\beta_{\Theta_\text{II},\text{i},\text{right}}).$

We proceed in a similar fashion for quantifying the reduction of the output bound.
In our setting with token-bucket arrival curves $\alpha$ and rate-latency service curves $\beta$, the output bound will be a token-bucket arrival curves with its rate equal to the bounded \ac{foi}'s rate.
Thus, we shift our attention to the output bound burstiness $(\alpha\oslash\beta)(0)$ that equals the \ac{foi}'s end-to-end backlog bound\footnote{Alternatively, the output burstiness can be computed as the bound on the \ac{foi}'s backlog in its last crossed server's queue \cite{Bondorf2016b}. This can improve results when \ac{FIFO} is not considered in the analysis but it remains unproven if this alternative derivation can reduce bounds in the analysis of \ac{FIFO} systems.} $\vdev$ (see \cref{thm:Performance-Bounds}).
After subtraction of the \ac{foi}'s inherent burstiness $b_{\text{foi}} = b$, we get the \textit{output (bound) burstiness increase} $\vdevinc$.
Its improvement is defined analog to Equation~14 as: 
\begin{eqnarray}
& & \hspace{-7.5mm} \frac{\mathit{output\,burstiness}^\text{FIFO} - \mathit{output\,burstiness}^\text{FIFO-FP\textsubscript{III}}}{\mathit{output\,burstiness}^\text{FIFO}} \nonumber \\
& & \hspace{-7.5mm} = \frac{(\alpha_\text{foi}\oslash\beta_{\Theta_\text{I},\text{i},\text{left}})(0) \wedge (\alpha_\text{foi}\oslash\beta_{\Theta_\text{II},\text{i},\text{left}})(0)
}{\mathit{output\,burstiness}^\text{FIFO}} \nonumber \\
& & \hspace{40mm} - \hspace{2mm}  \frac{(\alpha_\text{foi}\oslash\beta_{\Theta_\text{III},\text{iii}})(0)}{\mathit{output\,burstiness}^\text{FIFO}} \nonumber \\
& &\hspace{-7.5mm} = \frac{(\vdev(\alpha_\text{foi},\beta_{\Theta_{\text{I}\wedge\text{II}},\text{i}}) - b) - (\vdev(\alpha_\text{foi},\beta_{\Theta_\text{III},\text{iii}}) - b)}{\mathit{output\,burstiness}^\text{FIFO}} \nonumber \\
& &\hspace{-7.5mm} = \frac{\vdevinc(\alpha_\text{foi},\beta_{\Theta_{\text{I}\wedge\text{II}},\text{i}}) - \vdevinc(\alpha_\text{foi},\beta_{\Theta_\text{III},\text{iii}})}{\mathit{output\,burstiness}^\text{FIFO}}
\end{eqnarray}
where $\vdev(\alpha_\text{foi},\beta_{\Theta_{\text{I}\wedge\text{II}},\text{i}}) = \vdev(\alpha_\text{foi},\beta_{\Theta_\text{I},\text{i},\text{left}}) \wedge \vdev(\alpha_\text{foi},\beta_{\Theta_\text{II},\text{i},\text{right}})$.

\begin{figure}[h]
	\centering
	\subfigure[Delay bound improvements]{\label{fig:ludbff_fp:delay_improvements}
		\hspace{-3mm}%
		\includegraphics[width=\columnwidth]{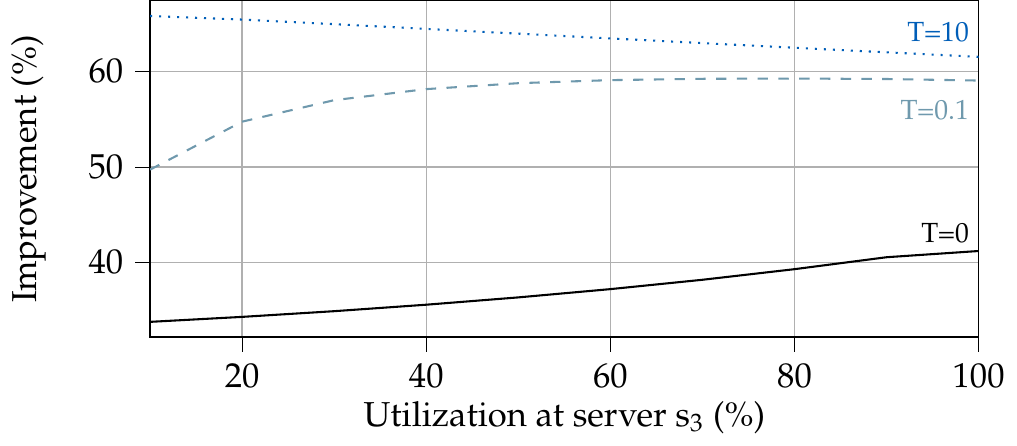}
	}
	\subfigure[Output bound burstiness improvement]{\label{fig:ludbff_fp:output_improvements}
		\hspace{-3mm}%
		\includegraphics[width=\columnwidth]{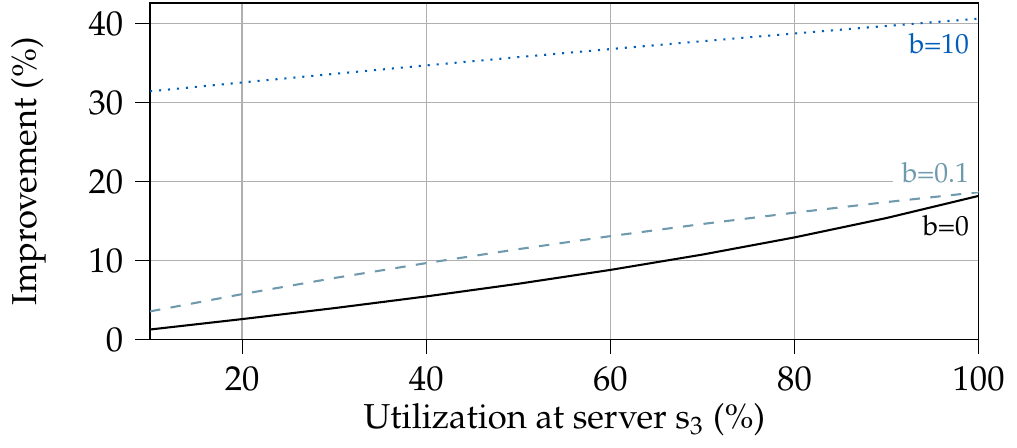}
	}
	\caption{Improvements of the \ac{foi}'s delay bound and output bound burstiness by applying \ac{FP} alternative \cref{fig:example_network:nc:single_fp} in the \ac{NC} \ac{FIFO} analysis of the network in \cref{fig:example_network:nc}}
	\label{fig:ludbff_fp_improvements}
\end{figure}

\Cref{fig:ludbff_fp_improvements} shows the results of evaluating these two improvements by \ac{FP}, subject to increasing peak utilization on the \ac{foi}'s path.
For both metrics, we achieve a considerable improvement already for lowest utilizations as well as $T=0$ and $b=0$, respectively.
These results motivate our work on integrating \ac{FP} into the feedforward analysis.

Notably, the delay bound improvement is influenced by the peak utilization by a far larger extent as shown in \cref{fig:ludbff_fp:delay_improvements}.
For $T=10$, we even observe a decrease of improvement.
An important impact factor is again the lack of the PMOO property.
In both \leftover service curves, $\beta_{\Theta_\text{I},\text{i},\text{left}}$ and $\beta_{\Theta_\text{II},\text{i},\text{right}}$, cutting causes the need to compute the output bound of the first subtandem, see occurrences of the deconvolution $\oslash$ in both terms.
If the operation takes a \leftover service curve, not only the latency $T$ is considered but also yet another latency increase.
This increase is grows with the peak utilization, not leading to a linear increase of the improvement.
We leave a thorough investigation for future work.

\subsection{The Challenge to Apply \acl{FP}}
\label{sec:fp4ludb:challenge}

The study of \ac{NC} analysis scalability demands appropriate tool support.
For the analysis of feedforward networks, we extended the implementation provided by the \ac{NCorgDNC} \DNCv \cite{Bondorf2014,Scheffler2021}.
Previous work \cite{GeyerSchefflerBondorf_RTAS2021} used the tool provided by Bisti et al.~\cite{Bisti2008,Bisti2012} that is limited to the study of tandem networks\footnote{http://cng1.iet.unipi.it/wiki/index.php/Deborah \cite{Bisti2010}}. 
Additionally, we benchmark against the \ac{FFMILPA} and \ac{FFLPA} \cite{Bouillard2015} by using another tool\footnote{\mbox{https://github.com/bocattelan/DiscoDNC-FIFO-Optimization-Extension}, \FFMILPAv. A tool for \cite{Bouillard2022} is not publicly available.}.
A recent overview on further \ac{NC} tools can be found in~\cite{Zhou2020}.

\subsubsection*{Analysis Scalability}
\label{sec:fp4ludb:scalability}

As mentioned in \cref{sec:introduction} and illustrated in Figures~\ref{fig:approach}(a) and \ref{fig:example_network:nc:all}, on each tandem with $n$ hops and $m$ cross-flows, \ac{FP} may explore $O(n^m)$ prolongation alternatives.
This raises doubts about its general scalability, confirmed in \cite{Bondorf2017c}.
Moreover, \ac{FFMILPA} is known to not scale well either.
Its heuristic \ac{FFLPA} is assumed to neither do so \cite{Bouillard2022}.
Therefore, we ran all analyses on a sample dataset of networks with up to 500 flows (see \cref{sec:gnn}), setting a time limit of \SI{1}{\hour} and a memory limit of  \SI{5}{GB} for each flow analysis.

\begin{figure}[h!]
	\centering
	\includegraphics[width=.85\columnwidth]{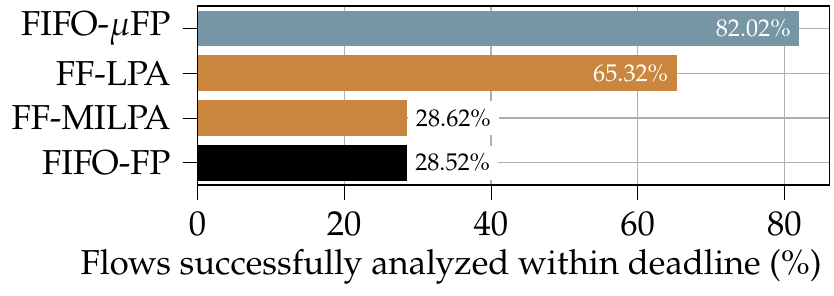}
	\caption{Flows successfully analyzed with a \SI{1}{\hour} deadline and at most \SI{5}{GB} memory usage}
	\label{fig:analysis_success}
\end{figure}

Results are presented in \cref{fig:analysis_success}.
\LUDBFFFP, the exhaustive search for the best \ac{FP} alternative on all tandems, was able to terminate for only \SI{28}{\percent} of flows. 
The successful termination ratio is roughly equal for \ac{FFMILPA}.
However, as both analyses are fundamentally different, their sets of analyzed flows are only equal for about \SI{57.5}{\percent}.

\subsubsection*{\acl{FP}'s Explored Alternatives}
\label{sec:fp4ludb:ncombinations}

The sheer amount of possible prolongation alternatives is the very cause of \LUDBFFFP's limited scalability.
In an attempt to scale \LUDBFFFP to analyze a larger set of flows, we created a heuristic that ignores certain alternatives.
We base this heuristic, called \LUDBFFhFP, on \cite{Bondorf2017c} where the analysis paired with \ac{FP} does not suffer from cutting and thus improvement \ref{itm:fp_improv:aggr} from above was evaluated in isolation.
It did not show significant potential.
In \LUDBFFhFP, we therefore restrict the search for a beneficial \ac{FP} alternative to check for improvement \ref{itm:fp_improv:cuts}.
We do so by checking for non-nested interference patterns and prolonging all involved flows to match the path of the longest flow.
\LUDBFFhFP may create new non-nested patterns by a prolongation, yet, we do not check any further.
\Cref{fig:analysis_success} already shows the amount of analyzed flows at \SI{82}{\percent} highest among all analyses.

\begin{figure}[h!]
	\centering
	\includegraphics[width=\columnwidth]{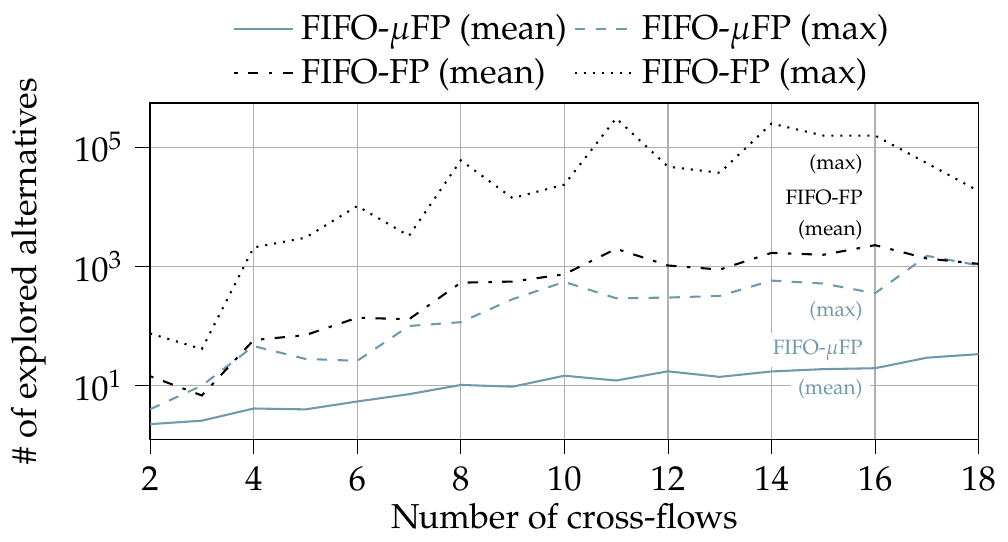}
	\caption{Relation between the number of cross-flows and the explored \ac{FP} alternatives by \LUDBFFhFP and \LUDBFFFP}
	\label{fig:ncrossflows_vs_ncombinations}
\end{figure}

Still, the amount of prolongation alternatives for the dataset evaluated in this article is forbiddingly large, see \cref{fig:ncrossflows_vs_ncombinations}.
Note for a large number of cross-flows, networks may have been excluded from this preliminary evaluation due to the limits set for computing data.
As expected, we get an exponential scaling between the number of cross-flows and the number of explored alternatives.

Last, to allow further fine-tuning of the tradeoff between delay bound tightness and computational effort, we may set the number of explored \ac{FP} alternatives to a fixed $k$ in our analyses. E.g., we will evaluate the performance of \MethodNameN{k} for some values $k\geq 1$ in \cref{sec:numerical_evaluation}.


\section{Effective \ac{FP} Predictions with \acsp{GNN}}
\label{sec:contribution}

We develop our universal \MethodName heuristic in this section, based in part on the work proposed in DeepTMA~\cite{Geyer2019a,Geyer2020}.
As illustrated in \cref{fig:approach}, the main intuition behind \MethodName is to avoid the exhaustive search for the best prolongation by limiting it to a few alternatives.
The heuristic's task is then only to predict the best flow prolongations, which are then fed to the \ac{NC} analysis.
This ensures that the bounds provided are formally valid.

For \MethodName, we used a \ac{GNN} as heuristic, since it was shown in DeepTMA to be a fast and efficient method \cite{Geyer2019a,Geyer2020b}.
We define networks to be in the \ac{NC} modeling domain and to consist of servers, crossed by flows.
We refer to the model used in \ac{GNN} as graphs.
During the different steps of the \ac{NC} \ac{FIFO} analysis, the sub-networks of interest are passed to the \ac{GNN} by transforming the networks into graphs (see \cref{sec:graph_transformation}) and processing them with the \ac{GNN}.
The outputs of the \ac{GNN} are then fed back to the \ac{NC} analysis, which finally performs its (min,plus) computations using the prolongations suggested by the \ac{GNN}.

\subsection{\acl{GNN} Fundamentals}
\label{sec:graphnn}


We use the framework of \acp{GNN} introduced in~\cite{Gori2005,Scarselli2009}.
They are a special class of \acp{NN} for processing graphs and predict values for nodes or edges depending on the connections between nodes and their properties.
The idea behind \acp{GNN} is called \emph{message passing}, where so-called \emph{messages}\ie vectors of numbers $\B{h}_v \in \R^k$, are iteratively updated and passed between neighboring nodes.
Those messages are propagated throughout the graph using multiple iterations until a fixed point is found or until a fixed number of iterations.
We refer to \cite{Gilmer2017} for a formalization of many concepts recently developed around \acp{GNN}.

Each node in the graph $v$ has input features\eg server rate, flow burstiness, labeled $\B{i}_v$, and output features\eg prolongation choices, labeled $\B{o}_v$.
Specific input and output features used for \MethodName will be explained in \Cref{sec:graph_transformation}.

The final messages are then used for predicting properties about nodes, namely the flow prolongations in our case.
This concept can be formalized as:
\begin{align}
	\B{h}_v^{(t)} & = \mathit{aggr}\left( \left\{ \B{h}_{u}^{(t-1)} \; \middle| \; u \in \textsc{Nbr}(v) \right\} \right) \label{eq:gnn_f_def} \\
	\B{o}_v       & = \mathit{out}\left( \B{h}_v^{(t \to \infty)} \right) \label{eq:gnn_output} \\
	\B{h}_v^{(t=0)} & = \mathit{init}\left(\B{i}_v \right) \label{eq:gnn_init}
\end{align}
with $\B{h}_v^{(t)}$ representing the message from node $v$ at iteration $t$, $\mathit{aggr}$ a function which aggregates the set of messages of the neighboring nodes $\textsc{Nbr}(v)$ of $v$, $\mathit{out}$ a function transforming the final messages to the target values, and $\mathit{init}$ a function for initializing the messages based on the nodes' input features.

Various approaches to GNNs have been recently proposed in the literature, mainly reusing the message passing framework from \Crefrange{eq:gnn_f_def}{eq:gnn_init} with different implementations for the $\mathit{aggr}$ and $\mathit{out}$ functions.
We selected \ac{GGNN}~\cite{Li2016a} for our \ac{GNN} model, with the addition of edge attention.
For each node $v$ in the graph, its message $\B{h}_v^{(t)}$ at iteration $t$ is updated at each iteration as:
\begin{align}
	\B{h}_v^{(t=0)} & = \mathit{FFNN}_{\mathit{init}}\left(\B{\mathit{inputs}}_v \right)  \label{eq:gnn_init} \\
	\B{h}_v^{(t)} &  = \mathit{GRU}\left( \B{h}_{v}^{(t-1)}, \sum_{u \in \textsc{Nbr}(v)} \lambda_{(u, v)}^{(t-1)} \B{h}_{u}^{(t-1)} \right) \label{eq:gnn_f_sum} \\
	\lambda_{(u, v)}^{(t)} & =  \sigma \left( \mathit{FFNN}_{\mathit{edge}}\left( \left\{ \B{h}_{u}^{(t)}, \B{h}_{v}^{(t)} \right\} \right) \right) \label{eq:edge_attention} \\
	\mathit{outputs} & = \mathit{FFNN}_{\mathit{out}}(\B{h}_v^{(d)}) \label{eq:gnn_output}
\end{align}
with $\sigma(x) = 1 / (1 + e^{-x})$ the sigmoid function,
$\textsc{Nbr}(v)$ the set of neighbors of node $v$,
$\{\cdot, \cdot\}$ the concatenation operator,
$\mathit{GRU}$ a \ac{GRU} cell, 
$\mathit{FFNN}_{\mathit{init}}$, $\mathit{FFNN}_{\mathit{edge}}$, $\mathit{FFNN}_{\mathit{out}}$ three \acp{FFNN},
and $\lambda_{(u, v)} \in (0,1)$ the weight for $(u, v)$.
The final prediction for each node is performed using \iac{FFNN} (see \cref{eq:gnn_output}) after applying \cref{eq:gnn_f_sum} for $d$ iterations, with $d$ corresponding to the diameter of the analyzed graph.

\subsection{Model Transformation}
\label{sec:graph_transformation}

Since we work with \iac{ML} method, we need an efficient data structure for describing \iac{NC} network which can be processed by \iac{NN}.
We chose undirected graphs, as they are a natural structure for describing networks and flows.
Due to their varying sizes, networks of any sizes may be analyzed using our method.

\begin{figure}[h!]
	\centering
	\hspace{12.5mm}\includegraphics[width=.725\columnwidth]{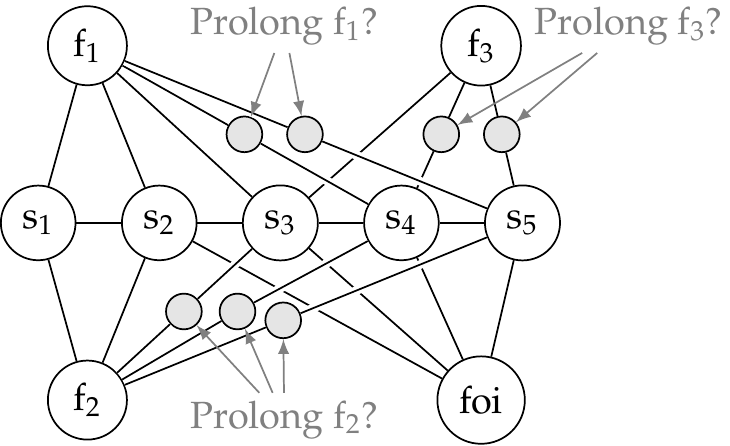}
	\caption{Graph encoding of the network from \cref{fig:example_network:nc}}
	\label{fig:example_network:graph}
\end{figure}

We only use the servers and flows relevant to the current \ac{foi} for the graph transformation by recursively backtracking through the server graph based on the \ac{foi}'s path.
We name this sub-network as \emph{sub-network of interest} $\Nfoi$.
We follow \cref{algo:graphtransformation} for this graph transformation, also illustrated and applied in \Cref{fig:example_network:graph} on the network from \cref{fig:example_network:nc}.
Each server is represented as a node in the graph, with edges corresponding to the network's links.
The features of a server node are its service curve parameters, namely its rate and latency, and its order w.r.t the \ac{foi}'s path.
Each flow is represented as a node in the graph, too.
The features of a flow node are its arrival curve parameters, namely its rate and burst in case of a token-bucket arrival curve.
Additionally, the \ac{foi} receives an extra feature representing the fact that it is the analyzed flow.

\begin{algorithm}[h!]
	\caption{Graph transformation of network $\Nfoi$}
	\label{algo:graphtransformation}
	\begin{algorithmic}
		\State $\G :=$ empty undirected graph
		\State \textbf{for all} server $s_i$ \textbf{in} network $\Nfoi$
			\textbf{do} $\G$.addNode($s_i$)
		\State \textbf{for all} link $(s_i, s_j)$ \textbf{in} network $\Nfoi$
			\textbf{do}  $\G$.addEdge($s_i$, $s_j$)
		\ForAll{flow $f_i$ \textbf{in} network $\Nfoi$}
			\State $\G$.addNode($f_i$)
			\State \textbf{for all} server $s_j$ \textbf{in} $f_i$.path()
				\textbf{do} $\G$.addEdge($f_i$, $s_j$)
		\EndFor
		\ForAll{flow $f_i$ \textbf{in} network $\Nfoi$ excluding \ac{foi}}
			\ForAll{server $s_j$ \textbf{in} \ac{foi}'s path}
				\If{prolongation $P_{f_i}^{s_j}$ of flow $f_i$ to $s_j$ is valid}
					\State $\G$.addNode($P_{f_i}^{s_j}$)
					\State $\G$.addEdges((${f_i}$, $P_{f_i}^{s_j}$), ($P_{f_i}^{s_j}$, $s_j$))
				\EndIf
			\EndFor
		\EndFor
		\Return $\G$
	\end{algorithmic}
\end{algorithm}

To encode the path taken by a flow in this graph, we use edges to connect the flow to the servers it traverses.
Compared to the original DeepTMA graph model from~\cite{Geyer2019a}, we simplify one aspect: we do not include path ordering nodes that tell us the order of servers on a crossed tandem.
The ordering is represented by the server order feature, which is the distance relative to the \ac{foi}'s sink.

To represent the flow prolongations, \emph{prolongation} nodes ($P_{f_i}^{s_j}$) connecting the cross-flows to their potential prolongation sinks are added to the graph.
Those nodes contain the hop count according to the \ac{foi}'s path as main feature~-- 
this is sufficient to later feed the prolongation into the \ac{NC} analysis, path ordering nodes are not required for this step either.
A prolongation node is also used for the last server of a cross-flow's unprolonged path (e.g. $s_4$ for $f_1$ in \cref{fig:example_network:graph}).
Those nodes represent the choice to not prolong a given flow.

Based on this graph representation, the task of the \ac{GNN} is to predict prolongation for each cross-flow by choosing the last server to prolong to.
Namely for each cross-flow $f$ and each potential sink $s$, the \ac{NN} assigns a probability value $\Pr_{f,s}$ between 0 and 1 to the corresponding prolongation node.
For each flow, the prolongation node with the highest probability decides which sink to use for prolonging the flow.
As illustrated in \cref{fig:approach}(b), those predictions are then fed to \LUDBFFFP, which finally performs the \ac{NC} analysis.

\subsection{Implementation} 
\label{sec:gnn}

We implemented the \ac{GNN} used in \MethodName using PyTorch~\cite{Paszke2019b} and \ac{PyG}~\cite{Fey2019}. 
Optimal parameters for the \ac{NN} size and the training were found using hyper-parameter optimization.
\Cref{tab:gnnsize} 
illustrates the size of the \ac{GNN} used for the evaluation in \cref{sec:numerical_evaluation}.

\begin{table}[h!]
	\centering
	\caption{Size of the \ac{GNN} used in \Cref{sec:numerical_evaluation}. Indexes represent respectively the weights ($w$) and biases ($b$) matrices}
	\label{tab:gnnsize}
	\resizebox{\hsize}{!}{%
		\begin{tabular}{ll}
			\toprule
			\textbf{Layer}                & \textbf{Size of the weights and bias matrices}                               \\ \midrule
			$\mathit{FFNN}_\mathit{init}$ & $(13, 128)_w + (128)_b$                                                      \\
			Message passing               & $(128, 128)_w + 2 \times \left\{(384, 128)_w + (128)_b \right\}$             \\
			$\mathit{FFNN}_\mathit{edge}$ & $\left\{(256, 128)_w + (128)_b\right\} + \left\{(128, 1)_w + (1)_b\right\}$  \\
			$\mathit{FFNN}_\mathit{out}$  & $\left\{(128, 128)_w + (128)_b \right\} + \left\{(128, 1)_w + (1)_b\right\}$ \\ \midrule
			\hfill Total:                 & \num{166402} parameters                                                      \\ \bottomrule
		\end{tabular}}
\end{table}

The \ac{GNN} and the \ac{NC} analysis are split into two processes during training, as shown in \cref{fig:training_approach}.
During evaluation, we integrated the graph transformation and the \ac{GNN} in \ac{NCorgDNC} using PyTorch's Java bindings, thus avoiding any inter-process communications.

\subsection{Dataset Generation}
\label{sec:dataset_generation}

To train and evaluate our \ac{NN} architecture, we generated a set of random topologies (as to check the \ac{FP} preconditions of \cref{sec:fp4ludb}) according to three different random topology generators:
\begin{enumerate*}[label=\textit{\alph*)}]
	\item tandems,
	\item trees and
	\item random server graphs following the $G(n,p)$ Erd{\H{o}}s-R{\'e}nyi model~\cite{Erdos1959}.
\end{enumerate*}
For each created server, a rate-latency service curve was generated with uniformly random rate and latency parameters.
A random number of flows was generated with random source and sink servers. 
Note that in our topologies, there cannot be cyclic dependency between the flows.
For each flow, a token-bucket arrival curve was generated with uniformly random burst and rate parameters.
All curve parameters were normalized to the $(0, 1]$ interval.

For each generated network, the \ac{NCorgDNC} \DNCv \cite{Bondorf2014} is then used for analyzing each flow.
We extended the feedforward \ac{LUDB} analysis of \cite{Scheffler2021} by the \ac{FP} feature to implement \LUDBFFFP, \LUDBFFhFP as well as \MethodName.
Each analysis is run with a maximum deadline of \SI{1}{hour} and maximum RAM usage of \SI{5}{GB}.

%
Since \LUDBFFFP may not bring any benefits compared to \LUDBFF, either due to no alternatives for prolonging flows or no end-to-end delay improvement by any alternative, we restrict the dataset to networks and flows where \ac{FP} is applicable\ie flows with prolongation options.
\Cref{tab:dataset_stats} contains statistics about the generated dataset.
In total approximately \num{50000} flows were generated and evaluated for the training dataset, and \num{20000} for the numerical evaluation presented in \cref{sec:numerical_evaluation}.
The evaluation dataset is split into two parts: small networks with up to 40 flows as in the training set, and larger networks with 100 to 493 flows.
The dataset will be available online to reproduce our results.

\begin{table}[h!]
	\centering
	\caption{Statistics about the generated dataset}
	\label{tab:dataset_stats}
	\begin{tabular}{l|rrr|rrr}
		\toprule
		\hfill \textbf{Dataset} &     \multicolumn{3}{c|}{\textbf{Train}}     &   \multicolumn{3}{c}{\textbf{Evaluation}}   \\ \midrule
		\textbf{Parameter}      & \textbf{Min} & \textbf{Mean} & \textbf{Max} & \textbf{Min} & \textbf{Mean} & \textbf{Max} \\ \midrule
		\# of servers           &            5 &          10.0 &           15 &            5 &          11.2 &           30 \\
		\# of flows             &           12 &          32.0 &           40 &            5 &         192.1 &          493 \\
		Flow path len           &            3 &           4.0 &            6 &            3 &           4.6 &           14 \\
		\# of cross-flows       &            6 &          20.7 &           31 &            2 &         132.4 &          492 \\ \bottomrule
	\end{tabular}
\end{table}

\subsection{\acl{NN} Training}
\label{sec:gnntraining}

We use \iac{RL} approach~\cite{Sutton2018}
for training the \ac{GNN}, where the optimal solution\ie the optimal flow prolongations, is not required to be known for training the \ac{GNN}.
This is different from previous works applying \acp{GNN} to \ac{NC}~\cite{Geyer2019a,Geyer2020b,GeyerSchefflerBondorf_RTAS2021} where \ac{SL} was used\ie where the best solution was known by running the exhaustive \ac{FP} analysis.

\subsubsection{\Acl{RL}}
\label{sec:gnntraining:rl}

\begin{figure}[!ht]
	\centering
	\includegraphics[width=\columnwidth]{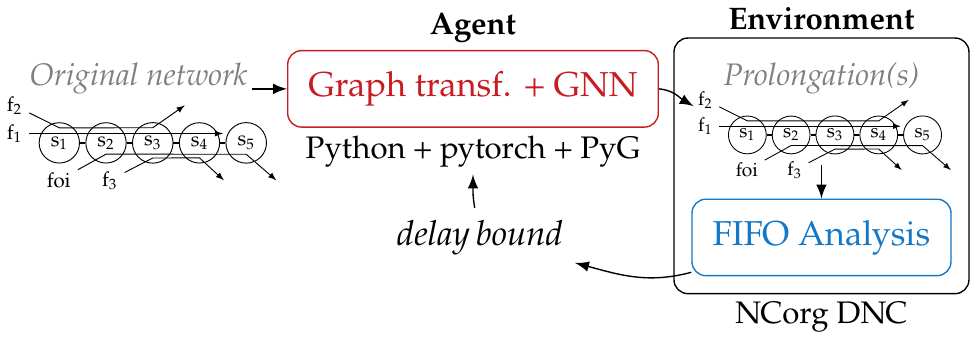}
	\caption{\ac{GNN} training using the \ac{RL} approach}
	\label{fig:training_approach}
\end{figure}

Our \ac{RL} approach is illustrated in \cref{fig:training_approach}.
The \ac{GNN} interacts with the \ac{NC} \ac{FIFO} analysis by providing it with flow prolongations.
As a feedback, the \ac{GNN} receives the end-to-end delay or output bound corresponding to the provided flow prolongations.
This bound is used for computing the reward, defined as the relative bound improvement over the standard \ac{FIFO} analysis without prolongations:
\begin{equation} \label{eq:reward_function}
	\reward = \frac{\mathit{bound}_{\text{FIFO}} - \mathit{bound}_{\text{FP}}}{\mathit{bound}_{\text{FIFO}}}
\end{equation}
The \ac{GNN} is trained to maximize the reward\ie produce prolongations leading to the best improvements over the standard \ac{FIFO} analysis.
Prolongations leading to a worse bound will result in a negative reward.

As opposed to more advanced applications of \ac{RL} where a series of actions are required, the agent produces here only a single action\ie the \ac{GNN} produces a single set of flow prolongations to apply to a given network and its flow(s) of interest.
Hence, we use here the REINFORCE algorithm \cite{Williams1992}, a policy gradient \ac{RL} approach.
The agent's decision making procedure is characterized by a policy $\pi(a,s,\omega) = \Pr(a|s,\omega)$, with $a$ the actions\ie flow prolongations, $s$ the state\ie the server graph, and $\omega$ the parameters of the policy\ie the weights of the \ac{GNN}.
The policy $\pi$ is defined as a categorical distribution for each flow, where the categories correspond to the different servers with the potential for prolongation.

At each training iteration, the actions $a$ are randomly sampled from the policy $\pi$ and passed to the \ac{NC} \ac{FIFO} analysis.
As feedback, the resulting delay bound from the \ac{NC} analysis is received and the reward is calculated according to \cref{eq:reward_function}.
The parameters of the policy $\pi$ are then updated as:
\begin{equation} \label{eq:policy_update}
	\omega \gets \omega - \nabla_\omega \log \pi(a,s,\omega) \cdot \reward
\end{equation}

To improve the training of the \ac{RL} policy, we used three well-known approaches.
First, we use an $\epsilon$-greedy exploration strategy where the action is either randomly sampled from the current policy with probability $1 - \epsilon$, or randomly sampled from a uniform distribution with probability $\epsilon$.
Secondly, we use \ac{UREX}~\cite{Nachum2017} to encourage undirected exploration of the reward landscape.
This adds a factor in \cref{eq:policy_update} which promotes actions with low probability from the current policy but high reward.
Our numerical evaluations showed that \ac{UREX} performed better than standard entropy regularization commonly used with REINFORCE.
Thus, we only use \ac{UREX} in the following.
Finally, we also use curriculum learning, where we gradually increase the difficulty of the prolongation task, numerically defined here as the product between the \ac{foi}'s path length and the number of cross-flows which can be prolonged.

\subsubsection{\Acl{SL}}

Our main motivation for using \ac{RL} is that \ac{SL} would require to extensively run the \ac{NC} \ac{FIFO} analysis on the training dataset for getting the optimal prolongations.
The computational cost of this task was already illustrated by the small percentage of analyses that finish within the execution time and memory restrictions (\cref{fig:analysis_success}).
To evaluate the impact of a limited dataset in \ac{SL} as compared to \ac{RL}, we also trained the \ac{GNN} using \iac{SL} approach.
We follow the method explained in \cref{sec:gnntraining:rl}, with action $a$ being inferred from the result of \LUDBFFFP or \LUDBFFhFP instead of sampling the policy $\pi$ for a random action.
We illustrate the impact of \ac{SL} on the bounds quality in \cref{sec:eval:rl_vs_supervised}.

\subsubsection{Training for Output Bounding}
\label{sec:gnntraining:output}

Setting the free $\theta$ parameters in the \ac{NC} \ac{FIFO} analysis' \leftover service curve term can vastly differ between bounding delay and output.
Since \ac{FP} is used when bounding the output of flows, too, we also query the \ac{GNN} for predicting prolongations under this objective.
We specifically trained a version of \MethodName where the \ac{NC} \ac{FIFO} analysis computes the output bound as the $\reward$.
A numerical evaluation showed no significant gains using this additional effort compared to training solely on the prolongations for delay bounding.
\Ie the cuts removed by \ac{FP} for best improvement of the delay bound are with very few exceptions also those to be removed for improving the output bound.
Hence, we restrict to training \ac{GNN} with the delay bound as $\reward$.

\subsubsection{From Add-on Feature to Full Analysis}

As noted in \cref{sec:fp_background}, \ac{FP} was originally designed as an add-on feature for \ac{NC} analyses.
Our preliminary investigation using \ac{SL}~\cite{GeyerSchefflerBondorf_RTAS2021} showed promising results such that we diverge from this previous view in our \ac{FIFO} analysis.
We promote \MethodName to a full analysis, meaning we only evaluate the $k$-predicted \ac{FP} alternatives.
Note, that the original paths might be predicted, too, and that any predicted prolongation will give a valid delay bound (see \cref{sec:fp_background}).

\section{Numerical Evaluation}
\label{sec:numerical_evaluation}

Our numerical evaluation aims to answer two questions: 
\begin{enumerate}
  \item How much delay bound improvement can \ac{FP} achieve?
  \item How do the analyses scale?
\end{enumerate}
In order to illustrate the benefits of \MethodName, we also provide comparisons against two analyses that select prolongation alternatives at random:
\begin{enumerate}[label=R\arabic*)]
  \item \RNDFP{k} selects from all possible \ac{FP} alternatives
  \item  \RNDhFP{k} adds some expert knowledge to select only from the \ac{FP} alternatives that are explored by \LUDBFFhFP
\end{enumerate}
As before, $k$ denotes the number evaluated alternatives such that $k\geq 1$.
We may omit $k$ if $k=1$.

In the following, we show details about \MethodName performance in terms of tightness as well as execution time.
Improvements in both will directly be applicable to and have an impact on any real-world application of the \ac{NC} methodology.
All evaluations presented here were done with the evaluation dataset described in \cref{sec:dataset_generation}.
Each analysis was limited to \SI{1}{\hour} execution time and \SI{5}{GB} memory.

\subsection{Delay Bound Improvements}
\label{sec:eval:delay_bound}

We compare in this section the gain in tightness for the delay bound compared against \LUDBFF.
To evaluate this gain, we use the delay bound gap metric for showing its increase:
\begin{equation} 
	\mathit{delay\, bound\, gap}_\text{foi} = \frac{delay_{\text{foi}}^{\text{FIFO}} - delay_{\text{foi}}^{\text{method}}}{delay_{\text{foi}}^{\text{FIFO}}}
\end{equation}

\subsubsection{\Acl{RL} vs. \Acl{SL}}
\label{sec:eval:rl_vs_supervised}

Previous works applying \acp{GNN} to \ac{NC}~\cite{Geyer2019a,Geyer2020b,GeyerSchefflerBondorf_RTAS2021} used \iac{SL}-based approach.
For a comparison to \ac{SL}, we ran \LUDBFFFP and \LUDBFFhFP on the fraction of our training dataset that finished analyzing under the time and memory restrictions (see \cref{sec:fp4ludb:challenge}).
While \LUDBFFhFP yields inferior prolongations, it creates a larger dataset to learn from than \LUDBFFFP does.
Prolongations from both analyses were then used as target for training two separate \acp{GNN}.
The \ac{RL}-approach for training the \ac{GNN} was implemented as described in \cref{sec:gnntraining}.

\begin{figure}[h!]
	\centering
	\includegraphics[width=.95\columnwidth]{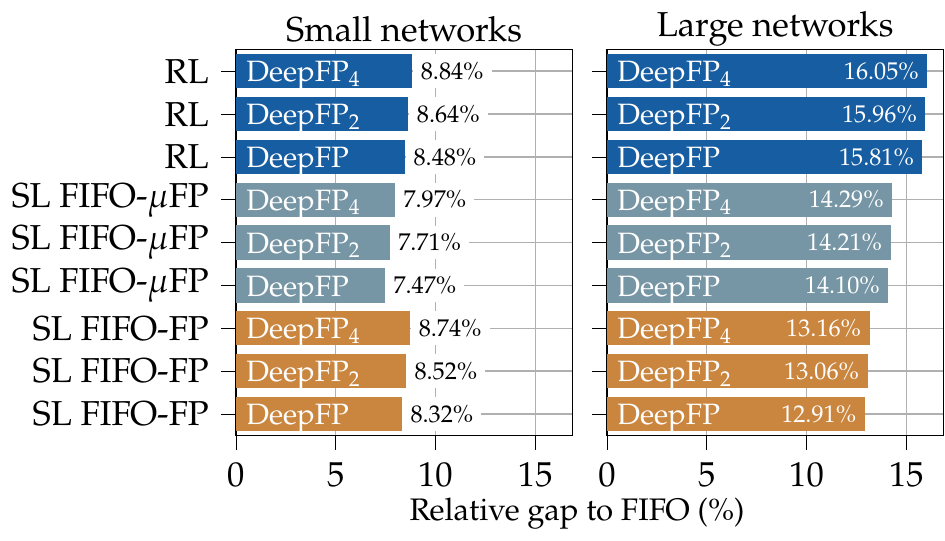}
	\caption{Impact of \ac{RL} vs. \ac{SL} on the average bound gap}
	\label{fig:reldelay_nofp_rl_vs_sl}
\end{figure}

On smaller networks, we notice that the predictions of the \acp{GNN} produce similar relative gaps when trained with \ac{RL} or \ac{SL} with \LUDBFFFP, with a small advantage for \ac{RL}.
On larger networks, \ac{RL} has a clear edge over \ac{SL}, stemming from two facts: 
\begin{enumerate*}[label=\textit{\alph*)}] 
	\item	 the training data for \ac{SL} could only be generated on a subset of the smaller networks and
	\item \ac{RL} can explore the aggregation potential of \ac{FP} (see \cref{sec:fp4ludb:ff_fp}).
\end{enumerate*}
This is further illustrated on the \ac{GNN} trained with \ac{SL} on \LUDBFFhFP, which achieves better delay bounds, yet still fails due to prolongations of lower quality than \LUDBFFFP.

Overall this illustrates that knowing the best prolongations is not necessarily required for training the \ac{GNN}.
Using a pure \ac{RL} approach without manually created training data can actually yield better results.

\subsubsection{\Acl{RL} performance}
\label{sec:eval:delay_bound:sl_v_rest}

In the following, we restrict our depiction on DeepFP using the \ac{RL}-trained \ac{GNN}.
As shown in \cref{sec:fp4ludb:challenge}, competing analyses were not necessarily able to analyze the entire evaluation dataset due to their poor scalability.
We restrict the following comparisons to those flows, across all considered networks, where the computation of the delay bound finished within \SI{1}{\hour} with maximally \SI{5}{GB} memory usage.
We present our results on increasing evaluation datasets.

\begin{figure}[h!]
	\centering
	\includegraphics[width=.825\columnwidth]{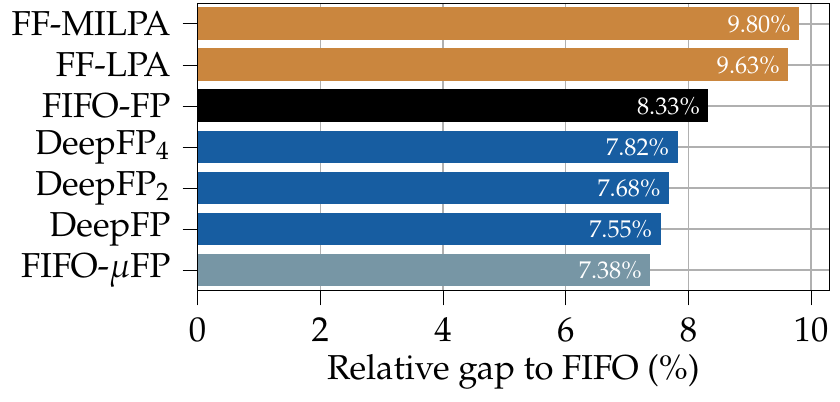}
	\vspace{-2.5mm}
	\caption{Average delay bound gap on the subset of flows successfully analyzed with \ac{FFMILPA}, \ac{FFLPA} and \LUDBFFFP}	\label{fig:reldelay_nofp_subset_fflpa}
\end{figure}

First, we compare \MethodName against the tight \ac{FFMILPA} and its heuristic \ac{FFLPA} on the flows that all analyses were able to analyze. 
\Cref{fig:reldelay_nofp_subset_fflpa} shows that \ac{FFMILPA} tightens delay bounds more than \LUDBFFFP and thus \MethodName.
However, this comes at a much greater computational cost as we show in \cref{sec:execution_time}.
Overall, \MethodName and \LUDBFFFP perform well and exploring multiple prolongation alternatives yields larger improvements yet these are only marginally better.

\begin{figure}[h!]
	\centering
	\includegraphics[width=.78\columnwidth]{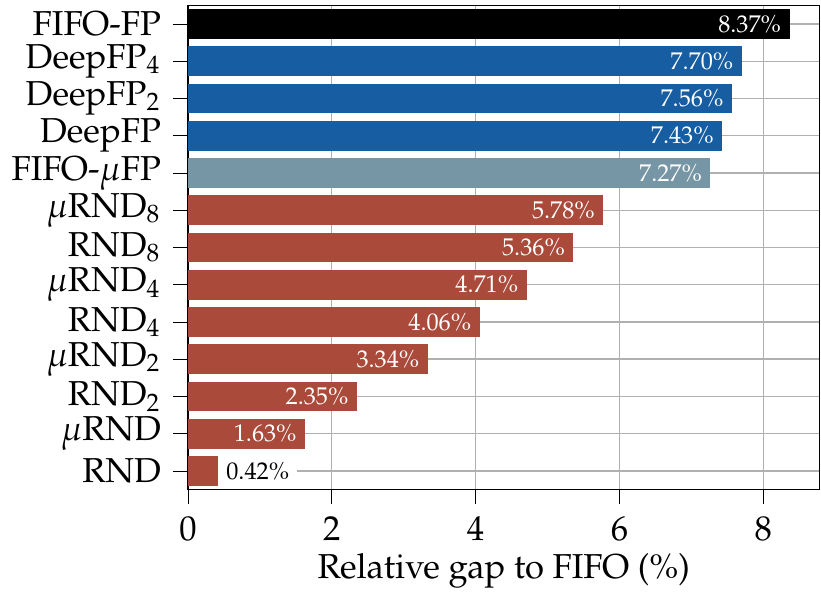}
	\vspace{-2.5mm}
	\caption{Average delay bound gap on the subset of flows successfully analyzed with \LUDBFFFP and \LUDBFFhFP}
	\label{fig:reldelay_nofp_subset_fullfp}
\end{figure}

An evaluation of the similarly large dataset that \LUDBFFFP was able to analyze with its naïve bruteforce approach is shown in \cref{fig:reldelay_nofp_subset_fullfp}.
It illustrates how well \MethodName performs compared to the best prolongations.
\MethodName computes on average similarly improved delay bounds as before, close to \LUDBFFFP.
Notably, on this dataset it beats the \LUDBFFhFP heuristic based on previous insights (see  \cref{sec:fp4ludb:ncombinations}).
Thus, it also beats \RNDhFP{k}, even with fewer explored prolongation alternatives.
\RNDFP{k} is less competitive than \RNDhFP{k}, as expected due to it selecting \ac{FP} alternatives from a larger pool that covers more potentially bad ones.

\begin{figure}[h!]
	\centering
	\subfigure[including comparison to \RNDFP{k}]{\label{fig:reldelay_nofp_subset_heuristicffp:rnd_fp}
		\includegraphics[width=.95\columnwidth]{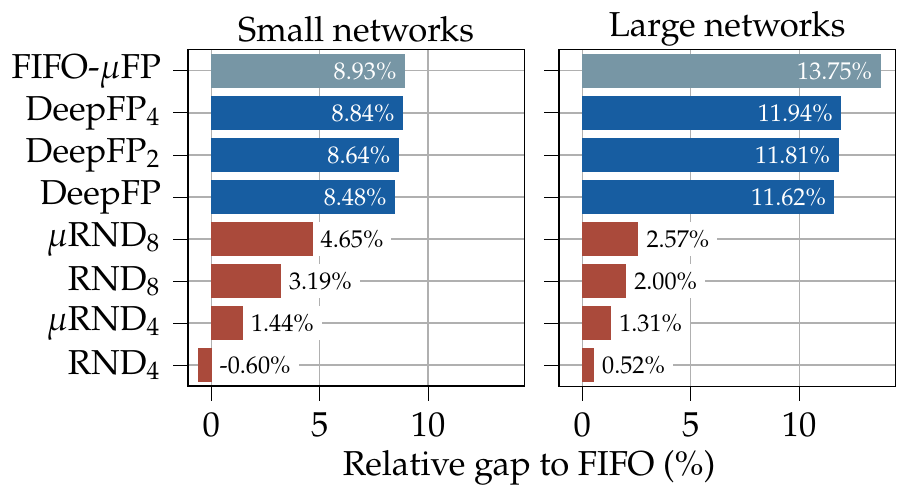}}
	\subfigure[including comparison to \RNDhFP{k}]{\label{fig:reldelay_nofp_subset_heuristicffp:rnd_h_fp}
		\includegraphics[width=.95\columnwidth]{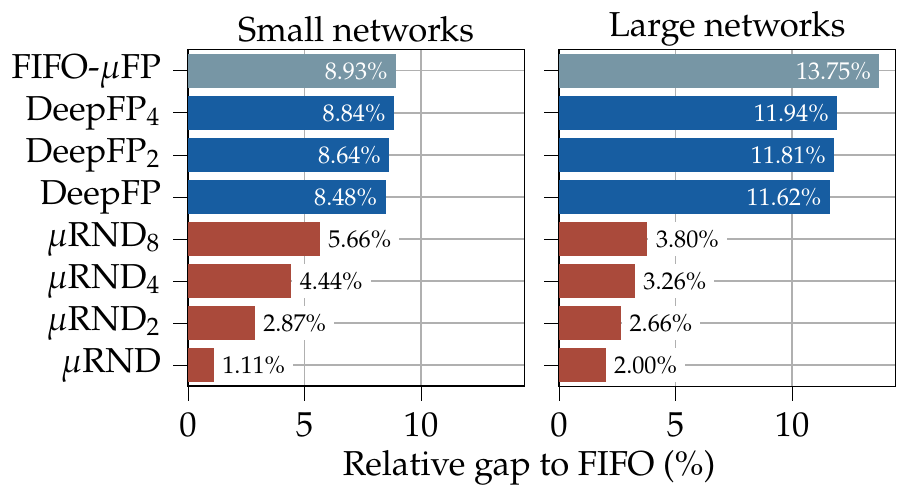}}
	\caption{Average delay bound gap on the subset of flows successfully analyzed with \LUDBFFhFP}	\label{fig:reldelay_nofp_subset_heuristicffp}
\end{figure}

Further increasing the dataset, we next compare \MethodName against \LUDBFFhFP in \cref{fig:reldelay_nofp_subset_heuristicffp}.
We partition networks into \textit{small} and \textit{large} to better visualize trends.
On the small networks, \MethodName's average delay bound gap is comparable with the one of \LUDBFFhFP but does not exceed it anymore.
The gap to \LUDBFFhFP increases on the large networks, indicating that \MethodName has some difficulties to generalize.

Moreover, the larger the networks become, the better \RNDhFP{k} seems to perform against \RNDFP{k}.
In the small networks in \cref{fig:reldelay_nofp_subset_heuristicffp}, we see a simple ranking by increasing $k$: \RNDFP{k} is outperformed by \RNDhFP{k} that, in turn, is outperformed by \RNDFP{k+1}. In the larger networks, however, \RNDhFP{} already performs as well as \RNDFP{4} and \RNDhFP{2} already outperforms \RNDFP{8}.
Still, the gap to \MethodNameN{} remains large for \RNDhFP{8}.

\begin{figure}[h!]
	\centering
	\includegraphics[width=.85\columnwidth]{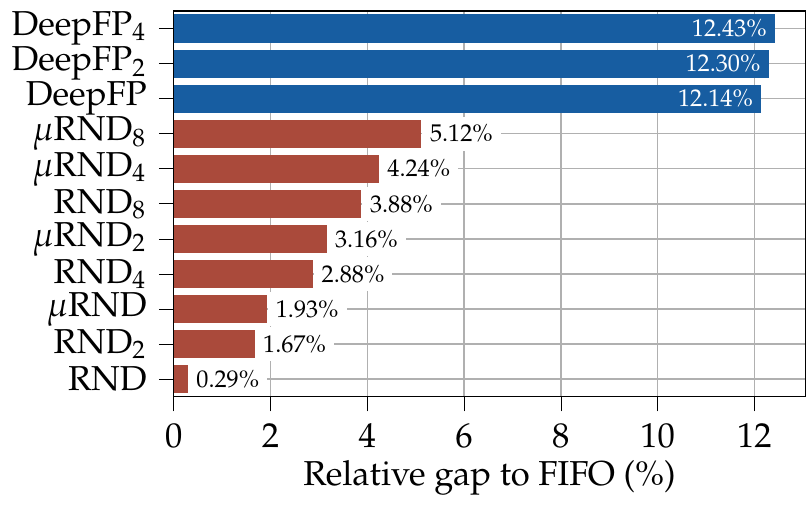}
	\caption{Average delay bound gap on the complete evaluation dataset}
	\label{fig:reldelay_nofp_fulldataset}
\end{figure}

Last, we compare \MethodName against \RNDFP{k} on the complete dataset in \cref{fig:reldelay_nofp_fulldataset}.
As expected, \MethodName is able to produce prolongations of much better quality than \RNDFP{k} and \RNDhFP{k}.
Even \MethodNameN{} outperforms \RNDhFP{8} by achieving more than double the relative gap to \LUDBFF.
Randomly choosing from the \LUDBFFhFP set of prolongations that was restricted to those alternatives deemed potentially beneficial by expert knowledge unfolds its potential here:
\RNDhFP{k} outperforms \RNDFP{2k}, i.e., we gain more improvement by lower effort.
More importantly, despite including tree and random networks in our evaluation, \MethodName retains the sizeable improvements obtained by the tandem-only analysis shown in \cite{GeyerSchefflerBondorf_RTAS2021}.

\subsubsection{Improvements by Increasing Training Effort}

\begin{figure}[h!]
	\centering
	\includegraphics[width=.95\columnwidth]{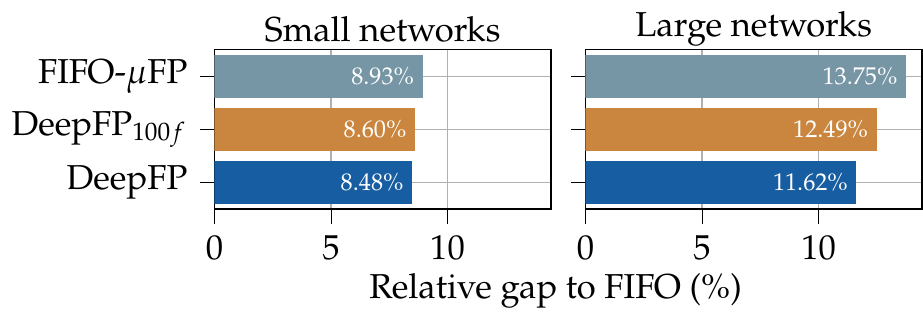}
	\caption{Average delay bound gap on the subset of flows successfully analyzed with \LUDBFFhFP}	\label{fig:reldelay_nofp_subset_heuristicffp_deepfpxl}
\end{figure}

We have put additional effort into training the \ac{GNN}.
We let the \ac{RL} approach learn in networks with up to 100 flows, compared to the maximum 40 flows of our previous evaluation.
We call the analysis benefiting from increased training effort \MethodNameN{$100f$}.
\Cref{fig:reldelay_nofp_subset_heuristicffp_deepfpxl} shows the results compared to \LUDBFFhFP on the dataset that could be analyzed with this analysis.
In the small networks, \MethodNameN{$100f$} closes the gap to \LUDBFFhFP but is not able to outperform it.
In the larger networks, this gap closing is more considerable.
Comparing \MethodNameN{$100f$} to \MethodNameN{$k$} in \cref{fig:reldelay_nofp_subset_heuristicffp}, we see that it performs similar to \MethodNameN{$2$} in small networks but can make use of increased training by even outperforming \MethodNameN{$4$} in the large networks where increasing $k$ did not yield significant improvements.

In the following evaluations, we will continue to consider \MethodNameN{}.

\subsubsection{Gap Distribution}

Our previous evaluations depict the average delay bound gap.
Last regarding the computation of delay bounds, we want to give some information about the distribution of the observed gaps to \LUDBFFhFP.

As \ac{FP} was shown to showed big potential to improve bounds, we decided to derive stand-alone analyses to benchmark on our dataset -- creating \MethodNameN{k}, \LUDBFFhFP, \RNDFP{k} and \RNDhFP{k}.
This is in contrast to previous work that interpreted \ac{FP} as a feature to add to the base analysis -- \LUDBFF in our case -- and eventually taking the smaller bound of both, \ac{FP} and non-\ac{FP} analysis.

\begin{figure}[h!]
	\centering
	\includegraphics[width=.85\columnwidth]{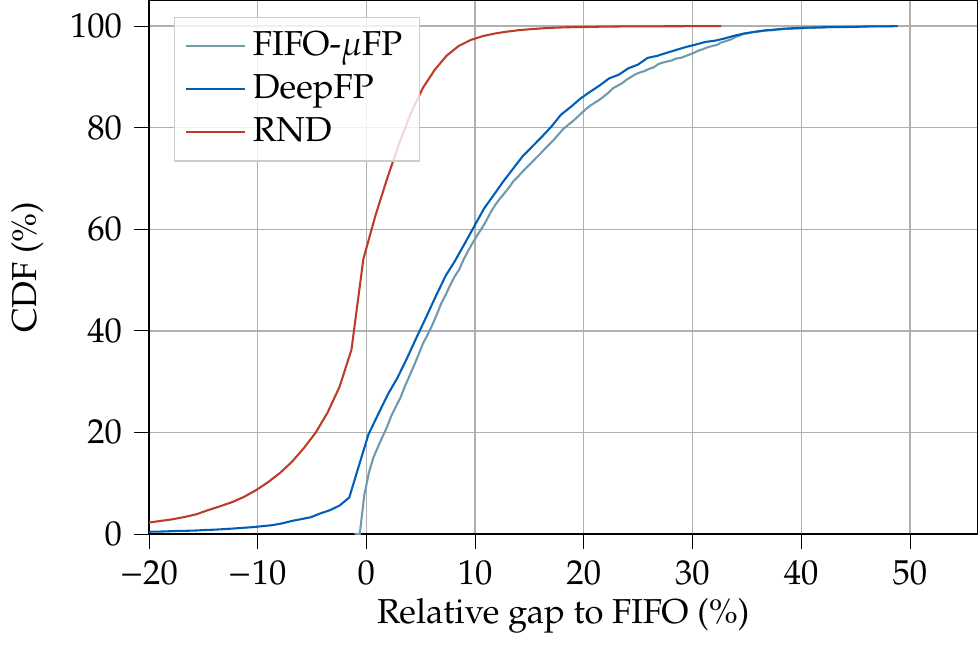}
	\caption{CDF of delay bound gaps on the subset of flows successfully analyzed with \LUDBFFhFP}	\label{fig:cdf_reldelay_nofp_subset_heuristicffp}
	\label{fig:cdf_reldelay_nofp_subset_heuristicffp:fp_stand_alone}
\end{figure}

Without the base analysis, we might feel the consequences of selecting bad prolongation alternatives.
This can already be seen for \RNDFP{} in \cref{fig:reldelay_nofp_subset_heuristicffp:rnd_fp} where a negative delay bound gap is ``achieved'' by \RNDFP{} (with $k=1$).
The delay bound gap distribution shown in \cref{fig:cdf_reldelay_nofp_subset_heuristicffp:fp_stand_alone} illustrates the reason\footnote{We omit \LUDBFFFP as its exhaustive enumeration covers the ``do not prolong any flow'' alternative, not giving insights in bad choices. Simultaneously, the effort of \LUDBFFFP analysis shrinks the analyzed dataset such that we would not be able to present meaningful results here.}: \RNDFP{} has a rather large share of flow analyses with negative gaps.
\MethodNameN{} and \LUDBFFhFP perform significantly better.
Yet, \MethodNameN{} still has some rather unfortunate choices whereas \LUDBFFhFP nearly eliminates them.
Both methods achieve a maximum gap of nearly \SI{50}{\percent}.



\subsection{Analysis Execution Times}
\label{sec:execution_time}

To understand the practical applicability of \MethodName, we evaluate in this section its execution time in different settings.
We define and measure the execution time per analysis as the total time taken to analyze each flow, including the time taken for initializing some data structures\eg graph transformation, conversion to \ac{LP}.
The execution times were measured on a server with Intel Xeon Gold 5120 CPU at \SI{2.20}{GHz} running Ubuntu 20.04 and OpenJDK \OpenJDKv with sequential garbage collection enabled.
IBM's CPLEX \CPLEXv was used as \ac{LP} solver\footnote{Notes on reproducibility of \ac{NC} results using IBM CPLEX as solver can be found in \cite{Scheffler2018}. Similar observations for the open-source alternative LpSolve are presented in \cite{Geyer2017b}.}.

Additionally, while part of the tools used here provide support for parallelization of some computations\eg CPLEX and PyTorch, these features were disabled and all evaluations were limited to a single CPU core by CPU pinning.
Furthermore, no GPU acceleration was used for the \ac{GNN}.
Neither was batching used\ie the \ac{GNN} analyzed one network at a time.
Overall, these settings ensure a fair evaluation and comparison between the different methods.

\begin{figure}[h!]
	\centering
	\includegraphics[width=\columnwidth]{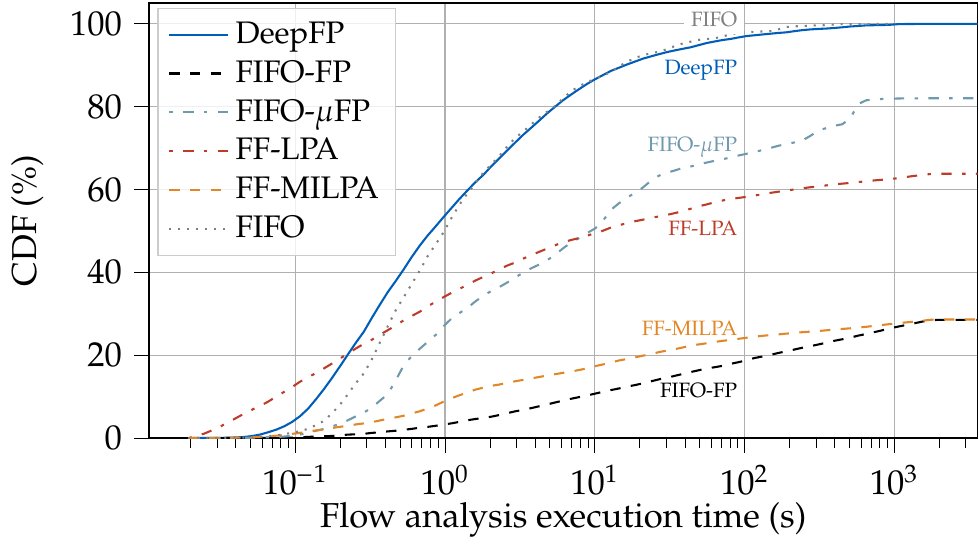}
	\caption{Total execution time of the different analyses}
	\label{fig:total_execution_time_w_fflpa}
\end{figure}

Measurements of the total execution time per flow analysis are summarized in \cref{fig:total_execution_time_w_fflpa}.
Overall, \MethodName exhibits execution times similar to \LUDBFF despite the overhead of the graph transformation and the \ac{GNN}.
It is even faster in some cases, which is explained by less decomposition steps thanks to the removal of cuts that also lead to a reduced amount of free $\theta$ parameters to optimize (see Equations~11 to \ref{eqn:beta_lo:fp_alt}) 
	as well as the complexity reduction due to \ac{FP} reported in \cite{Schmitt2007}.
	
Compared to \LUDBFFFP and \LUDBFFhFP, \MethodName is multiple orders of magnitude faster.
While \cref{fig:total_execution_time_w_fflpa} suggests that \LUDBFFFP and \ac{FFMILPA} as well as \LUDBFFhFP and \ac{FFLPA} roughly exhibit comparable execution times,
note that they do not analyze the same subset of flows under our restrictions.

\begin{figure}[h!]
	\centering
	\includegraphics[width=\columnwidth]{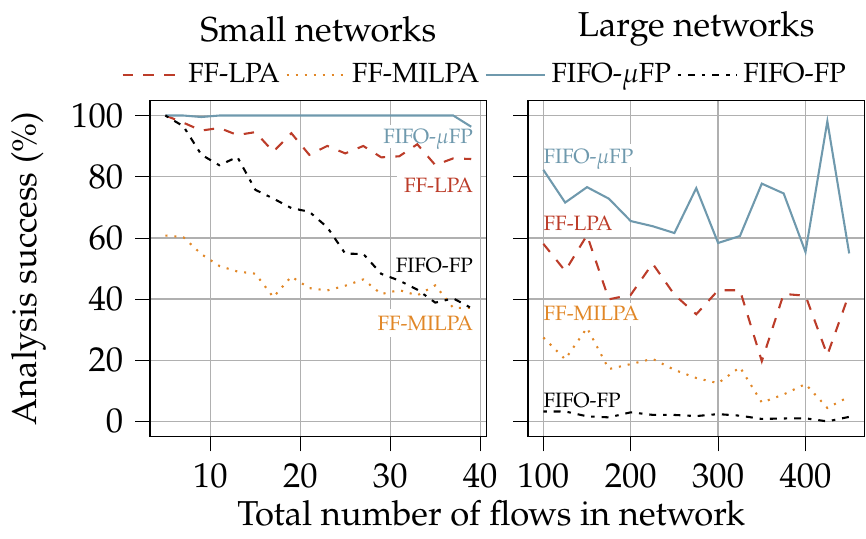}
	\caption{Analyses success per network size of flows with a \SI{1}{\hour} deadline and at most \SI{5}{GB} memory usage}
	\label{fig:analysis_success_detailed}
\end{figure}

For \LUDBFFFP, \LUDBFFhFP, \ac{FFLPA} and \ac{FFMILPA}, only a subset of the flows could be evaluated due to the \SI{1}{\hour} and \SI{5}{GB} memory cap per analysis as already shown in \cref{fig:analysis_success} and detailed in \cref{fig:analysis_success_detailed}.
This illustrates the poor scalability of these methods, even on networks of medium size.

\begin{figure}[h!]
	\centering
	\includegraphics[width=\columnwidth]{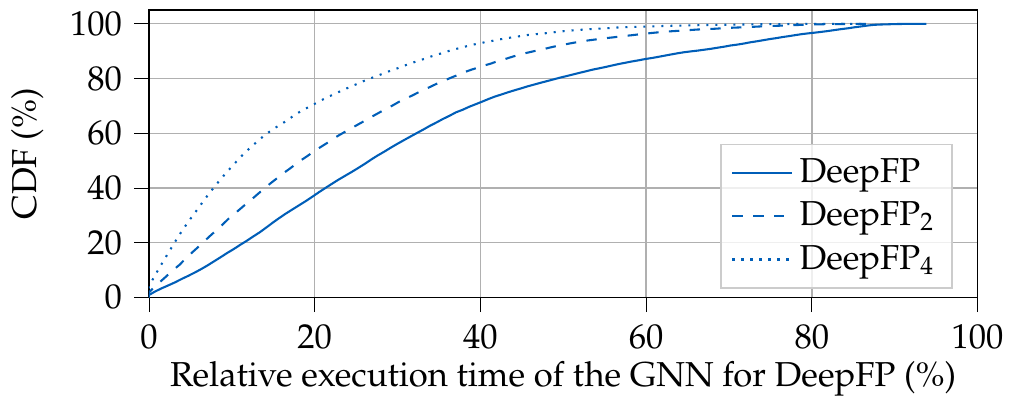}
	\caption{Relative execution time of the graph transformation and the \ac{GNN} in the \MethodName analysis}
	\label{fig:execution_time_gnn_share}
\end{figure}

Second, we evaluate the overhead of running the graph transformation and the \ac{GNN} and its execution time in comparison to the total execution time of the analysis.
We use the following measure:
\begin{equation}
	\frac{\mathit{Execution\,time\,GNN}}{\mathit{Total\,execution\,time}}
\end{equation}
Results are presented in \cref{fig:execution_time_gnn_share}.
In average, the \ac{GNN} requires \SI{31.8}{\percent} of the total execution time.
As illustrated earlier in \cref{fig:total_execution_time_w_fflpa}, this overhead has a low impact on the total execution time compared to the \LUDBFF analysis.

Compared to the earlier version of \MethodName from \cite{GeyerSchefflerBondorf_RTAS2021}, we use here a different integration of the \ac{GNN} in the \ac{NC} analysis.
We use PyTorch's Java API, reducing the overhead of inter-process communications initially required in \cite{GeyerSchefflerBondorf_RTAS2021}.

Overall, the improvements from \cref{sec:eval:delay_bound} and these execution time results illustrate that \MethodName is an effective method for tighter bounds at a low computational cost.

\subsection{Importance of \ac{NC} Parameters for \ac{GNN} Predictions}


For each parameter of the \ac{NC} network model presented in \cref{sec:gnn},
we aim to understand its importance for the \ac{GNN} to make a prediction.
We use the permutation-based importance measure \cite{Breiman2001} 
in order to assess this for \MethodName:
each feature is randomized by randomly permuting its values in the evaluation dataset, and assessing the impact on the delay bound gap $\mathit{delay\,bound\,gap}^\mathit{Feature}_\text{foi}$.
We define the \ac{GNN} input feature importance as:
\begin{equation}
	\mathit{delay\,bound\,gap}^\mathit{Baseline}_\text{foi} - \mathit{delay\,bound\,gap}^\mathit{Feature}_\text{foi}
\end{equation}
with $\mathit{delay\,bound\,gap}^\mathit{Baseline}_\text{foi}$ corresponding to the delay bound gap of \MethodName without column permutation.

\begin{figure}[h!]
	\centering
	\includegraphics[width=.8\columnwidth]{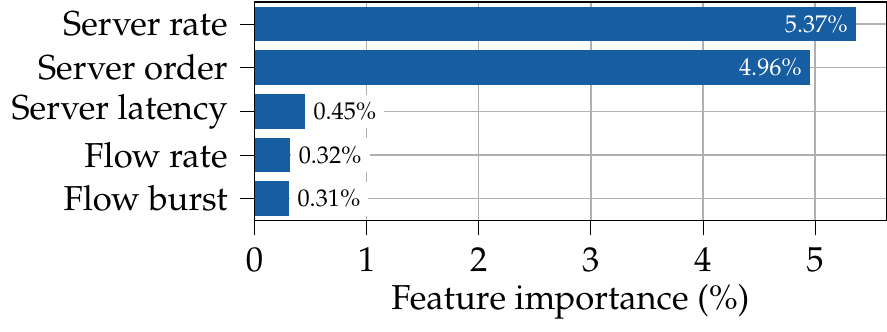}
	\caption{Feature importance of \MethodName}
	\label{fig:feature_importance}
\end{figure}

\Cref{fig:feature_importance} shows the results and confirms an expectation from the theory: the server rate is the feature having the largest impact on the prediction of the \ac{GNN}.
Prolongation is only possible if the servers to prolong over have sufficient spare capacity and trading off its disadvantage depends on the server rate.
Naturally, the server order has a similarly large impact on constructing new flow paths by prolongation.
The other features have almost two orders of magnitude less importance for the \ac{GNN} prediction compared to the server rate and order -- showing the \MethodName predictions' relative independence from them as from the actual bound to be computed, delay or output (see \cref{sec:gnntraining:output}).


\acresetall
\section{Conclusion}
\label{sec:conclusion}

In this article, we investigated \ac{FP} as a means to improve delay bounds of the algebraic \ac{NC} analysis for feedforward networks of \ac{FIFO} multiplexing and forwarding servers.
It is known that \ac{FP} does not scale.
We address this issue by a \ac{GNN} to create \MethodName, speeding up the \LUDBFF analysis with predictions while retaining validity of delay bounds.
Extending the initial work on this topic~\cite{GeyerSchefflerBondorf_RTAS2021}, we present a fast, tailored analysis that improves the bounds derived by non-\ac{FP} \LUDBFF analysis significantly.

We illustrate that the search for the best prolongations alternative is not computationally feasible for networks of even moderate size.
To overcome this exhaustive search, we propose in this article two approaches.
First, we introduce \LUDBFFhFP, a heuristic based on expert knowledge which considerably reduces the search-space.
Secondly, we use \MethodName, a heuristic based on \ac{GNN}, which learns to prolong flows using \iac{RL} approach.
With \ac{RL}, we contribute a novel training approach which directly interacts with the \ac{NC} \ac{FIFO} analysis and does not need the computationally expensive task of generating training labels required for \ac{SL}.
We show that this results in a better heuristic compared to a \ac{GNN} trained using \ac{SL}.

Via numerical evaluations, we show that both heuristics result in a reduction of the execution time by multiple orders of magnitude while still improving the delay bounds.
We also show that \MethodName is actually able to produce better prolongations than the expert-based heuristic on small networks.
Compared to another state-of-the-art based on \iac{LP} formulation, \MethodName is able to scale to much larger networks with only a small loss in tightness.

In conclusion, we show that \ac{FP} can considerably tighten \ac{NC} delay bounds for \ac{FIFO} feedforward networks and that the proposed \ac{GNN}-based \MethodName allows it to scale for application to larger networks.



{
\footnotesize
\yyyymmdddate
\bibliographystyle{IEEEtran}
\bibliography{IEEEabrv,biblio}
}

\end{document}